\documentclass[12pt]{article}
\usepackage{epsfig,graphicx,natbib,url,twoopt}
\usepackage[varg]{txfonts}
\usepackage{hyperref}          
\usepackage{pdfcomment,acronym}      
\usepackage{datetime}

\hypersetup{
  colorlinks=true,  
  urlcolor=blue,    
  linkcolor=red     
}
\usepackage[font={small,it}]{caption}

\def\RR #1RR{\noindent\small\textcolor{red}{$\sharp$RR: #1}\\[1ex]}
 
\bibpunct{(}{)}{;}{a}{}{,}    
\makeatletter
 \newcommandtwoopt{\citeads}[3][][]{%
   \nonstopmode
   \href{http://adsabs.harvard.edu/abs/#3}%
        {\def\hyper@linkstart##1##2{}%
         \let\hyper@linkend\@empty\citealp[#1][#2]{#3}}
   \biblink{#3}{\href{http://adsabs.harvard.edu/abs/#3}{ADS}}%
   \errorstopmode}            
 \newcommandtwoopt{\citepads}[3][][]{%
   \nonstopmode
   \href{http://adsabs.harvard.edu/abs/#3}%
        {\def\hyper@linkstart##1##2{}%
         \let\hyper@linkend\@empty\citep[#1][#2]{#3}}
   \biblink{#3}{\href{http://adsabs.harvard.edu/abs/#3}{ADS}}%
   \errorstopmode}            
 \newcommandtwoopt{\citetads}[3][][]{%
   \nonstopmode
   \href{http://adsabs.harvard.edu/abs/#3}%
        {\def\hyper@linkstart##1##2{}%
         \let\hyper@linkend\@empty\citet[#1][#2]{#3}}
   \biblink{#3}{\href{http://adsabs.harvard.edu/abs/#3}{ADS}}%
   \errorstopmode}            
 \newcommandtwoopt{\citeyearads}[3][][]{%
   \nonstopmode
   \href{http://adsabs.harvard.edu/abs/#3}%
        {\def\hyper@linkstart##1##2{}%
         \let\hyper@linkend\@empty\citeyear[#1][#2]{#3}}
   \biblink{#3}{\href{http://adsabs.harvard.edu/abs/#3}{ADS}}%
   \errorstopmode}            
\makeatother

\newcount\longrefs
\def\aap{\ifnum\longrefs=1 {Astron.\ Astrophys.}\else 
                           {A\hbox{\rm \&}A}\fi}
\def\aapr{\ifnum\longrefs=1 {Astron.\ Astrophys.\ Rev.}\else 
                            {A\hbox{\rm \&}AR}\fi}
\def\aaps{\ifnum\longrefs=1 {Astron.\ Astrophys.\ Suppl.}\else 
                            {A\hbox{\rm \&}A Suppl.}\fi}
\def\actaa{\ifnum\longrefs=1 {Acta Astronomica}\else
                            {Acta Astron.}\fi}
\def\aipcs{\ifnum\longrefs=1 {Am.\ Inst.\ Phys.\ Conf.\ Series}\else
                             {AIP Conf.\ Ser.}\fi}
\def\aj{\ifnum\longrefs=1 {Astron.\ J.}\else 
                          {AJ}\fi} 
\def\ao{\ifnum\longrefs=1 {Applied Optics}\else 
                           {Appl.\ Opt.}\fi} 
\def\aspcs{\ifnum\longrefs=1 {Astron.\ Soc.\ Pacific Conf.\ Series}\else 
                           {ASP Conf.\ Ser.}\fi} 
\def\apj{\ifnum\longrefs=1 {Astrophys.\ J.}\else 
                           {ApJ}\fi} 
\def\apjl{\ifnum\longrefs=1 {Astrophys.\ J. Lett.}\else 
                            {ApJL}\fi} 
\def\aplett{\ifnum\longrefs=1 {Astrophys.\ J. Lett.}\else 
                            {ApJ}\fi} 
\def\apjs{\ifnum\longrefs=1 {Astrophys.\ J. Suppl.}\else 
                            {ApJS}\fi}
\def\apss{\ifnum\longrefs=1 {Astrophys.\ and Space Science}\else 
                            {Astrophys.\ Space Sci.}\fi}
\def\araa{\ifnum\longrefs=1 {Ann.\ Rev.\ Astron.\ Astrophys.}\else 
                            {ARA\hbox{\rm \&}A}\fi}
\def\azh{\ifnum\longrefs=1 {Astronomicheskii Zhurnal}\else 
                            {Astron.\ Zhur.}\fi}
\def\baas{\ifnum\longrefs=1 {Bull.\ Am.\ Astron.\ Soc.}\else 
                            {BAAS}\fi}
\def\bain{\ifnum\longrefs=1 {Bull.\ Astronom.\ Institutes Netherlands}\else
                            {Bull.\ Astr.\ Inst.\ Neth.}\fi}
\def\cjaa{\ifnum\longrefs=1 {Chinese Jour.\ Astron.\ Astrophys.}\else 
                            {Chin.\ J.\ A\&A}\fi}
\def\gca{\ifnum\longrefs=1 {Geochim.\ Cosmochim.\ Acta}\else 
                           {Geochim.\ Cosmochim.\ Acta}\fi}
\def\grl{\ifnum\longrefs=1 {Geophys.\ Res.\ Lett.}\else 
                           {Geoph.\ Res.\ Lett.}\fi}
\def\iaucirc{\ifnum\longrefs=1 {IAU Circulars}\else 
                          {IAU Circ.}\fi}
\def\icarus{\ifnum\longrefs=1 {Icarus}\else 
                          {Icarus}\fi}
\def\ip{\ifnum\longrefs=1 {in press}\else 
                          {in press}\fi}
\def\jcap{\ifnum\longrefs=1 {Jour.\ Cosmology Astropart.\ Phys.}\else 
                          {JCAP}\fi}
\def\jgr{\ifnum\longrefs=1 {J.\ Geophys.\ Res.}\else 
                           {J.\ Geophys.\ Res.}\fi}  
\def\jrasc{\ifnum\longrefs=1 {J.\ Royal Astron.\ Soc.\ Canada}\else 
                           {JRAS Can.}\fi}  
\def\memsai{\ifnum\longrefs=1 {Mem.~Soc.~Astron.~Italiana}\else
                              {MmSAI}\fi}
\def\mnras{\ifnum\longrefs=1 {Mon.\ Not.\ Roy.\ Astron.\ Soc.}\else 
                             {MNRAS}\fi} 
\def\na{\ifnum\longrefs=1 {New Astronomy}\else 
                           {New Astron.}\fi}
\def\nar{\ifnum\longrefs=1 {New Astronomy rev.}\else 
                           {New Astron.\ Rev.}\fi}
\def\nat{\ifnum\longrefs=1 {Nature}\else 
                           {Nat}\fi}
\def\pasa{\ifnum\longrefs=1 {Pub.\ Astron.\ Soc.\ Australia}\else 
                            {PASA}\fi} 
\def\pasj{\ifnum\longrefs=1 {Pub.\ Astron.\ Soc.\ Japan}\else 
                            {PASJ}\fi} 
\def\pasp{\ifnum\longrefs=1 {Pub.\ Astron.\ Soc.\ Pacific}\else 
                            {PASP}\fi} 
\def\physscr{\ifnum\longrefs=1 {Physica Scripta}\else 
                            {Phys.\ Scrip.}\fi} 
\def\planss{\ifnum\longrefs=1 {Planetary \& Space Science}\else 
                            {Plan. \& Space Sci.}\fi} 
\def\procspie{\ifnum\longrefs=1 {Proc.\ SPIE}\else 
                            {Proc.\ SPIE}\fi} 
\def\qjras{\ifnum\longrefs=1 {Quarterly J.\ Royal Astron.\ Soc.}\else 
                            {QJRAS}\fi} 
\def\rmxaa{\ifnum\longrefs=1 {Revista Mexicana de Astron.\ y Astrofys.}\else 
                            {RMxAA}\fi} 
\def\sa{\ifnum\longrefs=1 {Soviet Astron..}\else 
                               {Sov.\ Astron.}\fi}
\def\skytel{\ifnum\longrefs=1 {Sky \& Telescope}\else 
                            {Sky \& Tel.}\fi} 
\def\solphys{\ifnum\longrefs=1 {Solar Phys.}\else 
                               {SoPh}\fi}
\def\sovast{\ifnum\longrefs=1 {Soviet Astronomy}\else 
                               {Sov.\ Ast.}\fi}
\def\ssr{\ifnum\longrefs=1 {Space Science Rev.}\else 
                               {Space\ Sci.\ Rev.}\fi}
\def\zap{\ifnum\longrefs=1 {Zeitschr.\ f.\ Astrophysik}\else
                               {Z.\ Astrophys.}\fi}

\makeatletter
\newcommand{\bibnote}[2]{\global\@namedef{#1note}{#2}}
\newcommand{\biblink}[2]{\global\@namedef{#1link}{#2}}
\makeatother

\newacro{AA}{Astronomy \& Astrophysics}  
\newacro{ADS}{Astrophysics Data System}
\newacro{AIA}{Atmospheric Imaging Assembly}
\newacro{ALMA}{Atacama Large Millimeter/submillimeter Array}
\newacro{AO}{adaptive optics}
\newacro{ApJ}{Astrophysical Journal}
\newacro{AR}{active region}
\newacro{bb}{bound-bound}
\newacro{bf}{bound-free}
\newacro{BFI}{Broad-band Filter Imager}
\newacro{CE}{coronal equilibrium}
\newacro{CfA}{Center for Astrophysics}
\newacro{CME}{coronal mass ejection}
\newacro{CRD}{complete redistribution}
\newacro{CRISP}{CRisp Imaging SpectroPolarimeter}
\newacro{CRISPEX}{CRisp SPectral EXplorer}
\newacro{CS}{coherent scattering}
\newacro{DEM}{Differential Emission Measure}
\newacro{DF}{dynamic fibril}
\newacro{DKIST}{Daniel K. Inouye Solar Telescope}
\newacro{DLR}{Deutsches Zentrum f\"ur Luft- und Raumfahrt}
\newacro{DOT}{Dutch Open Telescope}
\newacro{DST}{Richard B. Dunn Solar Telescope}   
\newacro{EB}{Ellerman bomb}
\newacro{EDP}{\'{E}dition Diffusion Presse}  
\newacro{EIT}{Extreme ultraviolet Imaging Telescope}
\newacro{EPIC}{European participation in Solar-C}
\newacro{ERC}{European Research Council}
\newacro{ESA}{European Space Agency}
\newacro{EST}{European Solar Telescope}
\newacro{EUV}{extreme ultraviolet}
\newacro{FAF}{flaring active-region fibril}
\newacro{ff}{free-free}
\newacro{FITS}{Flexible Image Transport System}
\newacro{FOV}{field of view}
\newacro{fov}{field of view}
\newacro{FWHM}{full width at half maximum}
\newacro{HAO}{High Altitude Observatory}
\newacro{HD}{hydrodynamics}
\newacro{Hi-C}{High Resolution Coronal Imager Sounding Rocket}
\newacro{HMI}{Helioseismic and Magnetic Imager}
\newacro{IAA}{Instituto de Astrof\'{i}sica de Andaluc\'{i}a}
\newacro{IAC}{Instituto de Astrof\'{i}sica de Canarias}
\newacro{IAS}{Institut d'Astrophysique Spatiale}
\newacro{IAU}{International Astronomical Union}
\newacro{IBIS}{Interferometric Bi-dimensional Spectrometer}
\newacro{IDL}{Interactive Data Language}
\newacro{IMaX}{Imaging Magnetograph eXperiment}
\newacro{INAF}{Istituto Nazionale di Astrofisica}
\newacro{IB}{IRIS bomb}
\newacro{IR}{infrared}
\newacro{IRIS}{Interface Region Imaging Spectrograph}
\newacro{ISAS}{Institute of Space and Astronautical Science}
\newacro{ISP}{Institute for Solar Physics}
\newacro{ISS}{International Space Station}
\newacro{ISSI}{International Space Science Institute}
\newacro{ITA}{Institute for Theoretical Astrophysics}
\newacro{JAXA}{Japan Aerospace Exploration Agency}
\newacro{JSOC}{Joint Science Operations Center}
\newacro{KIS}{Kiepenheuer--Institut f\"{u}r Sonnenphysik}
\newacro{KPNO}{Kitt Peak National Observatory}
\newacro{LASP}{Laboratory for Atmospheric and Space Physics}
\newacro{LC}{liquid cristal}
\newacro{LMSAL}{Lockheed Martin Solar and Astrophysics Labratory}
\newacro{LOS}{line of sight}
\newacro{LTE}{local thermodynamic equilibrium}
\newacro{MC}{magnetic concentration}
\newacro{MCAO}{multi-conjugate adaptive optics} 
\newacro{MDI}{Michelson Doppler Imager}
\newacro{ME}{Milne-Eddington} 
\newacro{MHD}{magnetohydrodynamics}
\newacro{MOMFBD}{Multi-Object Multi-Frame Blind Deconvolution}
\newacro{MPE}{Max--Planck--Institut f\"ur extraterrestrische Physik}
\newacro{MPG}{Max--Planck--Gesellschaft}
\newacro{MPS}{Max Planck Institute for Solar System Research}
\newacro{MSSL}{Mullard Space Science Laboratory}
\newacro{MTF}{modulation transfer function}
\newacro{NAOJ}{National Astronomical Observatory of Japan}
\newacro{NASA}{National Aeronautics and Space Administration}
\newacro{NIST}{National Institute of Standards and Technology}
\newacro{NLTE}{non-local thermodynamic equilibrium}
\newacro{NLFFF}{non-linear force-free field}
\newacro{NOAA}{National Oceanic and Atmospheric Administration}
\newacro{non-E}{non-equilibrium}
\newacro{NSO}{National Solar Observatory}
\newacro{NWO}{Netherlands Organisation for Scientific Research}
\newacro{PHE}{propagating heating event}
\newacro{PRD}{partial redistribution}
\newacro{PROBA2}{PRoject for OnBoard Autonomy}
\newacro{PSBE}{post Saha-Boltzmann extinction}
\newacro{PSF}{point spread function}
\newacro{QS}{quiet Sun}
\newacro{QSEB}{quiet-Sun Ellerman-like brightening} 
\newacro{RAL}{Rutherford Appleton Laboratory}
\newacro{RBE}{rapid blue-shifted excursion}
\newacro{R-MHD}{radiation hydrodynamics}
\newacro{rms}{root mean square}
\newacro{RMS}{root mean square}
\newacro{ROB}{Royal Observatory of Belgium}
\newacro{ROI}{region of interest}
\newacro{RRE}{rapid red-shifted excursion}
\newacro{RTE}{radiative transfer equation}
\newacro{RTSA}{Radiative Transfer in Stellar Atmospheres}
\newacro{SCF}{slender \CaIIH\ fibril}
\newacro{SE}{statistical equilibrium}
\newacro{SB}{Saha Boltzmann}
\newacro{SDO}{Solar Dynamics Observatory}
\newacro{SJI}{slit-jaw image}
\newacro{SLI}{slit image}
\newacro{SNR}{signal-to-noise ratio}
\newacro{SO}{Solar Orbiter}
\newacro{SoHO}{Solar and Heliospheric Observatory}
\newacro{SP}{Spectropolarimeter}
\newacro{SST}{Swedish 1-m Solar Telescope}
\newacro{SUMER}{Solar Ultraviolet Measurements of Emitted Radiation}
\newacro{SUFI}{Sunrise Filter Imager}
\newacro{SVD}{singular value decomposition}
\newacro{SVST}{Swedish Vacuum Solar Telescope}
\newacro{STX}{Solar Telescope X}
\newacro{THEMIS}{T\'{e}lescope H\'{e}liographique pour l'Etude du 
   Magn\'{e}tisme et des Instabilit\'{e} Solaires}     
\newacro{TR}{transition region}
\newacro{TRACE}{Transition Region and Coronal Explorer}
\newacro{TSI}{total solar irradiance}
\newacro{UT}{Universal Time}
\newacro{UV}{ultraviolet}
\newacro{VAULT}{Very high angular resolution ultraviolet telescope}
\newacro{VIRGO}{Variability of solar IRradiance and Gravity Oscillations}
\newacro{VTT}{Vacuum Tower Telescope}    
\newacro{XRT}{X-Ray Telescope}

\long\def\startignore #1\stopignore{}   

\def\rmit#1{{\it #1}}              
           
\def\ie{\rmit{i.e.,}}              
\def\eg{\rmit{e.g.,}}              

\def\specchar#1{\uppercase{#1}}    
\def\specand{ and }                
\def\specand{\,\&\,}               

\def\BaII{\mbox{Ba\,\specchar{ii}}}

\def\CaII{\mbox{Ca\,\specchar{ii}}}

\def\FeI{\mbox{Fe\,\specchar{i}}} 
\def\FeII{\mbox{Fe\,\specchar{ii}}}

\def\HeI{\mbox{He\,\specchar{i}}} 
\def\HeII{\mbox{He\,\specchar{ii}}}

\def\MgI{\mbox{Mg\,\specchar{i}}} 
\def\MgII{\mbox{Mg\,\specchar{ii}}}

\def\SrI{\mbox{Sr\,\specchar{i}}}

\def\Halpha{\mbox{H\hspace{0.1ex}$\alpha$}} 

\def\Lyalpha{\mbox{Ly$\hspace{0.2ex}\alpha$}}

\def\NaD{\mbox{Na\,\specchar{i}\,D}}    

\def\CaIIK{\mbox{Ca\,\specchar{ii}\,\,K}}       
\def\CaIIH{\mbox{Ca\,\specchar{ii}\,\,H}}
\def\CaIIHK{\mbox{Ca\,\specchar{ii}\,\,H{\specand}K}}
\def\HK{\mbox{H{\specand}K}}

\def\hk{\mbox{h{\specand}k}}

\def\level #1 #2#3#4{$#1 \; ^{#2} \mbox{#3} ^{#4}$}   

\def\kms{\hbox{km$\;$s$^{-1}$}}

\def\tis{\!=\!}                            

\font\dropfont= cmr12 scaled \magstep5
\def\dropcap#1#2{{\noindent
    \setbox0\hbox{\dropfont #1}\setbox1\hbox{#2}\setbox2\hbox{(}
    \count0=\ht0\advance\count0 by\dp0\count1\baselineskip
    \advance\count0 by-\ht1\advance\count0by\ht2
    \dimen1=.5ex\advance\count0by\dimen1\divide\count0 by\count1
    \advance\count0 by1\dimen0\wd0
    \advance\dimen0 by.25em\dimen1=\ht0\advance\dimen1 by-\ht1
    \global\hangindent\dimen0\global\hangafter-\count0
    \hskip-\dimen0\setbox0\hbox to\dimen0{\raise-\dimen1\box0\hss}
    \dp0=0in\ht0=0in\box0}#2}
\def\setlistparams{         
  \topsep=0.7ex                 
  \itemsep=0.7ex                
  \leftmargini=3ex}             
\setlistparams                  

\begin{document}  
\parindent=0ex
\parskip=1ex

\vspace*{-12ex}

{\em To appear in ``The National Solar Observatory at Sacramento Peak'',
editor S.E.~Keil}\\[2ex]

\begin{center}
{\Large \bf A YEAR AT SUNSPOT}\\[3ex]
{\Large Reminiscences of 1977--1978 at the Sacramento Peak Observatory}\\[3ex]
{\large Rob Rutten}\\[0.5ex]
 March 2018, Lingezicht Astrophysics, Deil, The Netherlands \\
\end{center}

{\em Forty years ago I spent a year with my family at the Sacramento
Peak Observatory in Sunspot, New Mexico. 
In August 2017 the U.S. National Solar Observatory ended its presence
there with a Farewell Workshop.
These reminiscences are in that context.
My aim is to paint the atmosphere}\footnote{Plus name dropping. 
Literally the Sac Peak atmosphere is most invigorating!
Cool, dust-free (but not pollen-free for those so afflicted), with a
fresh pine-tree aroma and pleasantly thin thanks to 2800~m altitude. 
Kim Streander who grew up as a Sac Peak kid would say ``When I get
back up there I can open my lungs and breathe, breathe!''. 
The only complainer was Dimitri Mihalas who moved from Boulder to Sac
Peak to escape Denver's pollution, but then saw to his dismay the
yellow plume from El Paso threatening him on the horizon -- 100\,km
away (at sunset one sees Gila ranges further away than the length of
Holland).
Back home I could initially play Bach sonatas with significantly fewer
breathing gasps, as the opening andante of the flute Partita (BWV
1013), a flow of semiquavers where any break is sin.
The best solar flutist is Andr\'as Ludm\'any at Debrecen. 
Three years later he composed a Bach sonata for me that was also too
hard for me, but he performed it later in a joint concert at Santa
Cruz de Tenerife.}.

\paragraph*{\dropcap{W}hen.}

I got my PhD at Utrecht University in the summer of 1976. 
My thesis adviser, Kees Zwaan\footnote{Zwaan was the last of the
former career pattern in which one became a high-school physics or
math teacher after leaving university with a ``doctorandus'' degree
(now Masters) and spent a decade of evening work to produce a PhD
thesis of sufficient quality to compete for the scarce professorships
-- or stay a teacher; many older teachers indeed held a PhD (nowadays
most are not academic). 
Zwaan so wrote a fat thesis earning him ``cum laude'' but he also
represented the transition by getting a grant from the Science
Foundation to complete the last part part-time at Sonnenborgh
Observatory.}, had spent a year at Sac Peak, loved it greatly (wore a
bolo tie ever since), and suggested I should go there rather than
Caltech/Big Bear or KPNO/Tucson as other major solar observatories.
I already held tenure\footnote{During a few years Kees de Jager, who
led the rapid expansion of Sonnenborgh Observatory during the generous
science funding in the 1960s after Minnaert's frugal reign, had a nice
habit of inviting young graduates to become staff member; this
happened to Henny Lamers, John Heise, Peter Hoyng and myself even
without applying for the job. 
This was the start of the new career pattern with funded
PhD-studentships, initially paid by the university, later also the
science foundation and nowadays the EU.  
Beforehand, Kees had already funded my graduation by appointing me and
another student as research assistants sharing this position's
funding, not telling us that we were essentially competing for the
full job and salary with our graduation research. 
Kees is presently 96, lives on the island of Texel, continues
publishing solar-cycle studies, and remains my life's example 
(apart from his running marathons until my present age;
km now).}; this was a post-PhD one-year outing to see solar
physics beyond Utrecht and hopefully obtain ideas and observations to
take home and never move again.

I applied for a Dutch Science Foundation fellowship to cover travel
and stay but didn't get it. 
However, in the interview chair J.H. Bannier\footnote{Ex-student of
Minnaert, then math teacher until he was picked as science
administrator after World War II, longtime director of the science
foundation, co-founder of CERN, ESO, ESRO\,$\Rightarrow$\,ESA. 
He didn't like my application statement that Minnaert's solar
spectrograph at Sonnenborgh was ripe for a museum, defining my need to
go abroad, and asked whether I thought that he belonged in a museum
too for having graduated with that instrument. 
Later it got a second life as a sun-as-star spectrograph
(\citeads{1982A&A...109...32O}) 
contributing to three more theses, but now it is indeed a museum
exhibit.} told me I might qualify for a NATO grant and asked whether I
had moral objections. 
Not me, hard to say yes for going to an USAF-founded and -funded
facility. 
Also, I had already unscrupulously used NATO money to travel through
Europe\footnote{After finding that NATO wanted conference organizers
to enlist young students, supposedly easier imprinted with NATO
doctrine. 
My brilliant study mate Hans Rosenberg and I so traveled to Greece and
Newcastle, where Chandrasekhar honored us in a dinner speech in the
great hall of Northumbria Castle as the only students amidst eight or
so Nobel winners.}
-- and did so later for the 1988 Capri solar-stellar granulation
workshop and the 1993 Soesterberg solar-stellar magnetism workshop. 
The Sac Peak Farewell workshop would have fit this milk-NATO strategy
well though not solar-stellar, but NATO is no longer into science.

The upshot is that we (me, Rietje -- Rita in the US -- and our
children Katrien and Martijn, then just 5 and nearly 4) arrived at
Sunspot in August 1977 (forty years before the Farewell) and returned
home in October 1978.

\paragraph*{\dropcap{F}irst impressions.}
George Simon had visited Utrecht just before.
He gave a Sonnenborgh colloquium on the five-minute oscillations, the
hottest solar physics topic at the time.
The unexpected highlight was that he also complained it was hot, asked
whether he might take off his jacket, then surprisingly bared himself
further taking off his tie and also his shirt  -- to show
the T-shirt underneath. 
It featured Jacques Beckers in massive curly hairdo.
I forgot George's excuse and context; of course Jacques came from
Utrecht but did his PhD work in Australia and I now wonder what the
intended message or pun or nastiness was\footnote{I later copied the
gimmick in an IAU meeting in Canada having the Sac Peak Tower on my
T-shirt; Jacques shouted from the back that I shouldn't show the
mirror -- seventy meters down in my crotch.}.

The next day we had George and Pat over for dinner and he drew us a
map of Alamogordo and the then Sunspot road,
instructing us to get a rental car at the airport and load up on
groceries at the Piggly Wiggly. 
Just the name! 
When we got there (flying the ``Vomit Comet'' from Albuquerque, indeed
a bumpy ride) we therefore did our first-ever shopping in America at
the Piggly Wiggly. 
I was flabbergasted to hear a most inviting ``Hello, how are you
today'' from the cashier girl; to me it sounded like ``please make
love to me'' which I didn't expect for my very first encounter with a
girl in the US.
I thought I knew American culture pretty well from reading Time
Magazine and John Updike and the like for years, but this was a
surprise. 

A yet more pleasant surprise was to meet a bunch of deer before the
Davis Ranch\footnote{Not yet appreciating that the humps besides the
road are former railroad grades, even put-aside ties still present.},
and then to find George Nelson and family awaiting us at the
observatory gate, the girls in pajamas while the rest of Sac Peak was
playing volleyball. 
Pinky had been a visiting student of Tony Hearn at Utrecht and was now
summer student with Jacques; we knew him well. 
He ran before our car to direct us to our relocatable
\#3069\footnote{In recent years Craig Gullixson's. 
The so-called relocatables were pre-fab houses, all identical with
three bedrooms and an open kitchen with two-way cupboards as divider,
a design we copied later in Lingezicht.} and then went on for
volleyball, making Gerda Beckers come over and say hello as
fellow-Dutch. 
This was an immediate introduction to the big-family style of life at
Sac Peak and the importance of volleyball there\footnote{I had at least
one summer student accepted by Steve Keil by selling him as good
volleyballer: Rieks Jager, later project manager for ALMA, ELT-METIS,
and more.}. 
We knew it was an isolated community but hadn't appreciated that
because of that the social fabric was very strong. 
We have never lived anywhere with so much interaction among neighbors
and the whole rest of the local community. 
America at its best in a most unlikely place\footnote{No other
observatory we visited had so much personnel living on the spot. 
At Kitt Peak people housed in Tucson, at NAOJ in surrounding Tokyo, at
Meudon in Paris suburbs, at Big Bear in Big Bear City.
At the Kiev Main Astronomical Observatory only a happy few had housing
on the beautiful observatory grounds. 
At the monumental Naples Osservatorio di Capodimonte only the
privileged director, as at Arcetri. 
At the Roque de los Muchachos the observers stay in the Residencia
while most telescope staff drives the mountain (except a few permitted
to sleep in the SST building, a historical exception there); we
enjoyed spending DOT summers in a cabin besides a deep barranco but an
hour drive or bike down and isolated. 
At Hida Observatory and Sayan Observatory we found similar
isolated-site companionship but with too small permanent populations.
At Sunspot during our year Jack Evans, Jacques Beckers and George
Simon had already moved to Alamogordo, Dick Dunn to Las Cruces, Lou
Gilliam and Christy Ott to Cloudcroft, but I guess we still amounted
to 80--100 (in 16 redwoods, 20 relocatables, 2 trailers, 7 BOQ
apartments). 
Later the population declined but the start of Apache Point
Observatory added some (including Bruce Gillespie who earlier took the
fabulous \CaIIK\ spectroheliogram on the proceedings of Bruce Lites'
1984 workshop). 
The close-knit but sizable Sunspot community has been very special.
}.

\paragraph*{\dropcap{W}ork.}
First I had to revamp my Utrecht Algol codes into Fortran. 
There were two computers, a Xerox Sigma\,5 in the Mainlab filling its
basement and a Sigma\,3 in the Tower. 
Both read laboriously punched cards\footnote{But there and at Utrecht
the card punchers, big noisy apparatus, printed the content of the
80-column card along the top (still familiar from 80-character fits
headers; I still set emacs to 80-character lines for my programs).
During my two-month stay at Kiev two years later during the bleak
Brezhnef stagnation I found that their main computer, an illicit
IBM-360 clone, could competitively run Fortran codes including Natasha
Shchukina's iterative two-level approximation NLTE program, the same
approach as Avrett's Pandora invented independently, but also that the
Kiev punch machine did not print what got punched.
Natasha used to quickly scan through the card pile (in foot-long
boxes) flipping each up against the light to detect a typo or find the
location to replace or insert cards by reading the eight-bit encoding
holes in each column as characters, but I didn't master that art and
quit programming there. 
Her Lena (aka Khomenko, then 3) probably played with punch cards; my
desk still rests on stacks of them applied as floor equalizer.
Earlier Utrecht University had relied on punched paper tape with eight
holes per character; 255 = erase was used to mask errors by inserting
a row of eight holes manually with a stylus and a tape-holder with
guiding holes (we also had fancy tape winders, even powered ones). 
Replacing a subroutine was done by cutting that segment out of a tape
(one typed programs with stretches of blank tape between subroutines)
and splice the new one in its place.    
Then there came the battle between the mathematicians and physicist
Tini Veltman whether to adhere to Algol'60 and paper tape or go modern
by moving to Fortran and punch cards.
Luckily Veltman won -- we got the CDC\,6600 he knew from CERN where he
developed ``schoonschip'', the engine for his Nobel-winning
electro-weak renormalization with my co-student Gerard 't Hooft.},
spit out lineprinter sheets in giant stacks, and filled magnetic reel
tapes. 
The Sigma\,5 even had huge multi-platter disk packs storing as much as
100 megabytes!  
One was reserved by the manager, Charles (Chuck) Bridges, for his
butterfly catalogs, his main passion apart from nightly running the
Cow Trail down to the West Side Road and back up for exercise. 
The Sigma~5 was booked per job for half or full hours, butterflies
took most nights. 
There was a monitor near the entrance door to the Mainlab showing who
had it; when my name was up Jacques would shout ``Another rotten
job!'', but most of the daytime went to Franz Deubner
Fourier-transforming microphotometered Tower spectra to measure
differentially-rotating deep-bound $p$-modes. 

Yes, spectrograms were on film! 
A fantastic detector re pixels.
Dick Dunn's Echelle Spectrograph ESG fed spectra over a full meter
length to 70-mm film at a frame rate of five seconds cadence: at 100
linepairs/mm resolution a 4\,K$\times$4\,K detector, like SDO's but
faster. 
Better yet was the ingenious multiplexing scheme with a pre-disperser
prism spectrograph feeding a slit plane in which multiple slits could
be moved (by shifting cover-up plates held by clamps or magnets) to
select different pieces of spectrum from different overlapping orders
(not spread out transversely as common for nighttime \'echelle
spectrographs but all overlapping each other at varying dispersion and
shift) to get to the film, with wide-band filters mounted above the
film to select the proper segment of the desired order\footnote{I once
spent half an hour folded up in the closed ESG box to get my eyes dark
adapted for precisely positioning a polarizer on a large-Land\'e blend
between \HK\ -- hard to see since so very violet. 
I might do better now, my current cataract-replacing plastic eye
lenses have higher ultraviolet transmission. 
Horst let me out on me tapping the V-sign he knew from having been on the
bad side -- deep into Russia which scared him enough to build a
cold-war atomic-bomb shelter for \#3049 on Hound Dog Hill on the
suspicion Holloman was a primary ICBM target.}.
This ingenious scheme permitted to make lines of interest from all
over the spectrum lie close together but still separate across the
film, so that one didn't register long swaths of non-interesting
continuum but just desired spectral segments with interesting lines,
at 70-mm film extent along the slit for spatial resolution.
There was a program to compute layouts for any set of desired lines. 
Jacques Beckers had the most comprehensive setup called HIRKHAD,
combining \CaIIHK, the \CaII\,IR triplet, \Halpha, the \NaD\ lines,
and more, I think also a Zeeman-sensitive \FeI\ line and a
non-Zeeman-sensitive \FeI\ velocity line.
Nowadays you may get such combinations at the SST combining CRISP and
CHROMIS (but I would like to add \BaII\,4554, \HeI\,D3, \HeI\,10830).

At that time the only means for digital recording were large unwieldy
single-pixel photomultiplier tubes in Peltier cooling boxes (I used
one at the 1970 Mexico eclipse as real-time exposure meter in the
spectrograph just before less wieldy though-the-lens metering became
an asset of Asahi Pentax Spotmatic cameras, and a moving one at the
Utrecht solar spectrometer to scan the spectrum digitally onto punched
paper tape). 
However, Dick Dunn was already pioneering 1D solid-state detection
with his Diode Array, still very large in itself to accommodate 512
pin diodes and enclosed in a large steel box mounted off the ESG on
the table edge.
The time of mounting spectrographs etc.\ as individual optical
components on human-sized benches came only with the small size of CCD
pixels. 
At that time there wasn't even much room for optical benches because
the Sigma 3 with its Hollerith punch-card reader, line printer and
many cabinets occupied much of the co-rotating telescope table.

The efficient Tower team led by Horst Mauter (with Dick Mann and Gary
Willis rotating on a 3-day late, 3-day early, 3-day off schedule)
would have your miles of film developed and ready for science the next
day. 
The snag was that they then had to get microphotometered; even Dunn's
``fast microphotometer'' next to the Sigma\,5 and noisily zipping
boustrophedonically over the film couldn't keep up. 
At Utrecht I had been a microphotometer buff myself, digitizing a
Houtgast-built one\footnote{Houtgast was the colorful second to
Minnaert in the 1930s. 
During World War II he replaced Minnaert who was interned by the
Germans with other leading Dutch scientists, and then permitted Kees
de Jager to hide from the Germans in the observatory.
His brilliant PhD thesis (\citeads{1942PhDT........12H}, my birth
year) settled the open issue whether one should account for coherence
in scattering; the answer that complete redistribution suffices for
most lines has been gospel since, with partial redistribution limited
to only a few lines (\CaII\,\HK, \MgII\,\hk, \Lyalpha\ -- and
\BaII\,4554\,\AA) and a few characters (including Han Uitenbroek and
me).
After the war Houtgast moved into a personal niche of eclipse
expeditions (I think because Kees de Jager rather than he was selected
by Minnaert as successor), taking me on three that defined my career. 
Earlier he equipped the microphotometer used to produce the Utrecht
Atlas of \citetads{1940pass.book.....M}, 
for which Mulders had taken the plates at Mount Wilson before becoming
science director at the NSF, with very clever analog computing to
convert the measured plate transmission into incident intensity by
letting a slit-formatted galvanometer beam scan the non-linear
calibration curve that was determined for each plate and then cut out
from cardboard using nail scissors.
The design is shown in the Atlas prefaces, in English but also in
Esperanto in keeping with Minnaert's idealism.}
to spit out punched paper tape at 100 samples/s, but when I returned
home with many large cans of ESG films and smaller cans of
corresponding 35-mm slitjaw films I had no clue how to handle them --
they rusted away. 
Like almost all other film cans still locked away for posterity (if
ever) in the plate vault in the Mainlab basement.

The same happened with the early-generation floppy\footnote{These were
still indeed floppy (bendable).} disks from the PDP-8 computer at the
Big Dome from a project using Jack Evans' double-pass spectrograph to
register the full \HK\ wings center-to-limb including absolute
calibration of the ``flux'' spectrum using Jacques' 5-mm cylindrical
sun-as-a-star telescope feed\footnote{The ``world's shortest and most
astigmatic solar telescope'' with which Jacques and Lou Gilliam
registered the flux atlas (\citeads{1976hrsa.book.....B}), by far the
handiest and most useful of all printed solar atlases. 
It has convenient letter size, not the unwieldy large formats of the
earlier Utrecht Atlas and the later Jungfraujoch and FTS atlases, and
all line identifications from the table by
\citetads{1966sst..book.....M} added alongside with a program by Chuck
Bridges. 
I used to take my much-annotated copy wherever I went observing and
also still have a Kurucz-generated digital version of it.  
Before that, somebody at Sac Peak (female I suspect) had been given
the laborious job of writing all identifications along the line dips
in a copy of the Utrecht Atlas that laid ready for inspection besides
the ESG spectrograph in the Tower, an example that Kees Zwaan followed
at Utrecht when he returned from his year at Sac Peak by putting his
personal computer Ed van der Zalm, my later coworker and friend, to
the same task; that annotated copy is now on display in the present
Sonnenborgh Museum as a gift from me. 
I didn't know Kees got the idea at Sac Peak until I saw the ESG copy;
Dick Mann commented that Kees should have asked a copy instead -- he
had it on microfilm and could easily produce another print with their
spectrum-printing machine.
At Utrecht personnel as Ed hired to do tabular calculations were then
called ``computer'', as in the acknowledgment in the
\citetads{1966sst..book.....M} line table. 
In the late sixties there were half a dozen at work at Sonnenborgh,
mostly manually converting spectrum plate densities into intensities
and variable-star measurements into light curves. 
At that time at Sac Peak Jack Evans would hire ``the girls'' -- the
astronomer's wives -- for such jobs: lady computers as were
Annie Cannon at Harvard and Annie Maunder at Greenwich, of better fame
than Joseph Stalin whose only employment was as computer at Tiflis
Observatory.}.
having Lou Gilliam spent a night with a standard lamp in a convenient
nearby tree (I believe to monitor coelostat-induced polarization
affecting the grating transmission), and the same happened with the
magnetic reel tape from the last week of our US caper when I
registered the full \HK\ profiles at super spectral resolution with
the Fourier Transform Spectrometer at Kitt Peak (still the bluest on
offer in the FTS database), also center-to-limb. 
The upshot of all this, I am ashamed to say, is that although I like
observing and am not bad at it, the subsequent data reduction is far
beyond me so that my track record in publishing results from my own
data is ashamedly lousy.  

Fortunately, there were other ESG films at Sac Peak that turned out
paper-producing, all taken by Jacques Beckers. 
They weren't even microphotometered but just printed by photographer
Dick Faller to prepare the publication figures.
The first came to my office as a 70-mm film in the hands of Lawrence
Cram who had found it in the large film cabinet in the Tower where he
regularly sat inspecting spectra per film viewer\footnote{A two-spool
device with between the spools a meter-long back-lit bed for the film
to slide through with a large magnifier mounted in the middle; the
best ones accommodated the 70-mm spectra film and both 35-mm slitjaw
films in parallel for simultaneous and time-delay image comparisons. 
I had one such constructed in Utrecht after my return and wonder what
happened to it. 
Recently I wrote an IDL many-movie viewer for similar inspection and
time-delay blinking (showex.pro in my imagelib) -- business as usual.}
that had been taken by others. 
He asked how a little \FeII\ line could appear in emission at the
center of the disk. 
It led to \citetads{1980ApJ...241..374C} 
showing that the little line responds to radiation from deeper even
than its background continuum (the inner \CaIIH\ wing).
Bruce Lites got involved while he was visiting to observe, I think
with David Elmore's Stokes-2 at the Big Dome, since he was willing to
set up a suitable NLTE \FeII\ computation using his SOSO version of
the Auer-Mihalas complete linearization code on the Cray supercomputer
at Boulder. 
We visited him there for this work, starting a life-long friendship. 

A similar full-linearization setup for Ba\,II was done for me by Bob
Milkey at Tucson whom we also visited a few times\footnote{During one
of those visits Karen Harvey kindly refilled her swimming pool just
for our kids; they had already been cheated out of the Alamogordo
swimming pool where water below 25 centigrade was judged unfit for
humans. 
They adamantly convinced Karen that they loved cold water. 
Later in Boulder they were accepted without problem into the deep CU
Olympic pool even though they couldn't swim. 
No swimming wheresoever at Sac Peak was the largest minus of the site
for them; they tried a pool off Bluff Springs but came out with too
many leeches. 
Unfortunately we didn't yet know the Scott Able Canyon creek on the
way to Timberon that probably was still year-long running at that
time, very scenic with beautiful and children-ready travertine slides. 
Now dry alas since the water drain for Orogrande (the much-ticketed
slow-down hamlet on the 54 from El Paso) was moved to the upper
creek.}
to confirm the empirical evidence of partial redistribution that I had
found in my thesis work
(\citeads{1979ApJ...231..277R}). 

The second set of Beckers spectra were brought by Bob Stencel, also
visiting (we reciprocally visited at Goddard on our way back), who had
asked Jacques for good limb spectra and got his best.
They covered the extended \CaII\ \HK\ wings with a curved slit just
inside the limb and with remarkable spatial resolution thanks to
superb seeing that people sort of jealously attributed to Jacques as a
special streak of his\footnote{Indeed I had my best seeing that year
on a day when I fell in for him because he had to stay in Alamogordo
for jury duty -- proud of being US citizen.
I suspect that the Tower might have been situated better just south of
the Big Dome, more like the Sloan telescope at the very edge of the
steepest south-facing slope, or indeed on Apache Point. 
Similarly, I think that the SST (and the DOT) should have been put on
La Palma's Pico de la Nieve on the east rim of the Caldera, having
much better upslopes for the prevailing north-east winds to bring
laminar conditions and push the caldera chimney plume from solar wall
heating away from the line of sight until late afternoon.
Nowadays one should mount automated Beckers SHABARs
(\citeads{2010SPIE.7733E.144S}) at many locations and also on quads
and do precise micro-comparisons.}.
Bob's prints of these spectra led to
\citetads{1980A&AS...39..415R}, 
an undercited publication also treating occurrence of emission before
the regular onset of the flash spectrum that I think still poses
questions such as the strange behavior of the Y\,II limb emission
lines. 
But also the other publications didn't make it into Hirsch-index ones;
only later did I improve my standing\footnote{Never suffering
``publish or perish'' permitted me to develop at my own slow speed. 
My highest hirscher is one of my only two publications addressing
other stars,
\citetads{1994A&A...288..860C}, 
addressing lithium NLTE in other stars.
While it had a negative result and remains the most boring analysis we
ever did, it remains my best cited -- many more astronomers are into
galactic abundances than into solar spectrometry: leave solar if you
wanna hirsch.}. 

Before these director Jack Zirker had complained about my lack of
productivity, not directly to me but via Jacques and also Tony Hearn
when he visited, triggering a comment from Lawrence Cram (who was
writing important publications at the time using HIRKHAD spectra): ``How can
he get something important if he just mucks around with Ba\,II and
Y\,II lines?''.
Indeed I was. 
Later on I did somewhat better by moving up in the periodic system,
first to \FeI\ with Bruce Lites' brilliant NLTE thesis as inspiration,
then to \CaII\ \HK\ and their grains, then the alkalis, then to
\CaII\,8542 and finally all the way up to \Halpha\ which at last I
think I understand (although it made me a terrorist
suspect\footnote{Lesson 1: don't forget your laptop in a plane. 
Lesson 2: don't leave an open email on your screen. 
Lesson 3: don't display emails about bombs. 
Even if these are just Ellerman bombs of 1917 vintage (well --
Ellermans release more energy than the champion 50-megaton Tsar bomb
of 1961, also a hydrogen one).
Lesson 4: the Dutch border police took three months to identify me as
Ellerman bomber, don't fear them if you are a bad guy.
Lesson 5: the NSA didn't blacklist me yet, the world ain't too bad
yet. 
But maybe these very sentences will do me in per Google Scholar.
Lesson 6: don't try to compete with Sami Solanki. 
During the Farewell I hoped to score the 100th citation of
\citetads{1917ApJ....46..298E} 
in its centennial year (its citation score shot up dramatically the
last years, thanks to ADS and partly thanks to me, implying that your
nice discoveries do get credit eventually -- just sit back and wait a
hundred years!) --  
but it looks like the centennial 100th citation was
\citetads{2017A&A...597A.127B} 
including ubiquitous Sami.}), and with \HeI\ and \HeII\ as tempting
goal on my horizon -- presently I work on Tom Schad multi-slit
10830\,\AA\ imaging spectrometry from the DST, a spin-off of the
Farewell. 

But so I always remained spectrally oriented and always remained
within the photosphere and chromosphere, whereas my contemporaries of
that time ranged far wider. 
Exemplary in the table below are Tim Brown, who went to other stars
and then to planets and became Sac Peak's highest hirscher, and Jeff
Linsky who after his thesis with Gene Avrett on \HK\ moved to Boulder
and other stars.

\begin{table}[htbp]
\caption[]{\label{tab:hirsch}
Publication tallies collected from ADS in January 2018. 
Left: Sac Peak staff and summer students during 1977--1978.
Right: Sac Peak visitors during 1977--1978.
Rich Robinson and Jack Thomas have ADS-confusing namesakes. }     
\begin{center}    
  \vspace{-2ex} 
  \begin{tabular}{|l|c|c|c|}
  \hline
  & pubs & Hirsch & ratio \\
  \hline
   Dick Altrock      & 158 & 17 & 0.11 \\
   Jacques Beckers   & 326 & 33 & 0.10 \\
   Tim Brown         & 529 & 76 & 0.14 \\
   Lawrence Cram     & 228 & 34 & 0.15 \\
   Franz Deubner     & 169 & 23 & 0.14 \\
   Dick Dunn         & 112 & 14 & 0.13 \\
   Jack Evans        &  61 & 13 & 0.21 \\
   Harry Jones       & 143 & 18 & 0.13 \\
   Steve Keil        & 125 & 18 & 0.14 \\
   Don Neidig        &  73 & 16 & 0.22 \\
   Steve Musman      &  55 & 14 & 0.25 \\
   Rich Robinson     &  ?  & plm 34 & ?    \\
   Rob Rutten        & 227 & 32 & 0.14 \\
   Tim Schneeberger  &  46 & 10 & 0.22 \\
   George Simon      & 111 & 23 & 0.21 \\
   Ray Smartt        & 145 & 15 & 0.10 \\
   Tuck Stebbins     & 95  & 16 & 0.17 \\
   Pete Worden       & 195 & 30 & 0.15 \\
   Jack Zirker       & 175 & 22 & 0.13 \\[1.0ex]
   Mark Giampapa    & 271 & 39 & 0.14 \\ 
   Jeff Kuhn        & 254 & 27 & 0.11 \\
   George Nelson    &  28 & 11 & 0.39 \\
   \hline
  \end{tabular}
\quad
  \begin{tabular}{|l|c|c|c|}
  \hline
  & pubs & Hirsch & ratio \\
  \hline
   Martin Altschuler & 45 & 16 & 0.22 \\
   Grant Athay      & 226 & 37 & 0.16 \\
   Larry Auer       & 126 & 34 & 0.27 \\
   Bart Bok         & 262 & 21 & 0.08 \\
   Dick Canfield    & 363 & 43 & 0.12 \\
   Stirling Colgate & 265 & 35 & 0.13 \\   
   Eric Fossat      & 223 & 32 & 0.14 \\
   Phil Goode       & 351 & 43 & 0.12 \\
   Tony Hearn       &  68 & 13 & 0.19 \\
   Henry Hill       & 135 & 12 & 0.09 \\
   Joe Hollweg      & 228 & 52 & 0.23 \\
   Stuart Jordan    &  76 & 11 & 0.15 \\
   Henny Lamers     & 452 & 60 & 0.13 \\ 
   Jeff Linsky      & $>$1000  & 76 & $>$ 0.08 \\
   Bruce Lites      & 292 & 55 & 0.19 \\
   Dermott Mullan   & 324 & 33 & 0.10 \\
   Gene Parker      & 479 & 69 & 0.14 \\
   Henk Spruit      & 301 & 61 & 0.20 \\
   Bob Stencel      & 484 & 35 & 0.07 \\
   Jan Stenflo      & 346 & 46 & 0.13 \\
   Dick Thomas      & 242 & 28 & 0.12 \\
   Jack Thomas      & ?   & plm 33 & ? \\  
   Charles Wolff    & 73  & 18 & 0.25   \\
   \hline
  \end{tabular}
\end{center}
\end{table}

\begin{figure}[hbtp]
  \centering
  \includegraphics[width=0.5\textwidth]{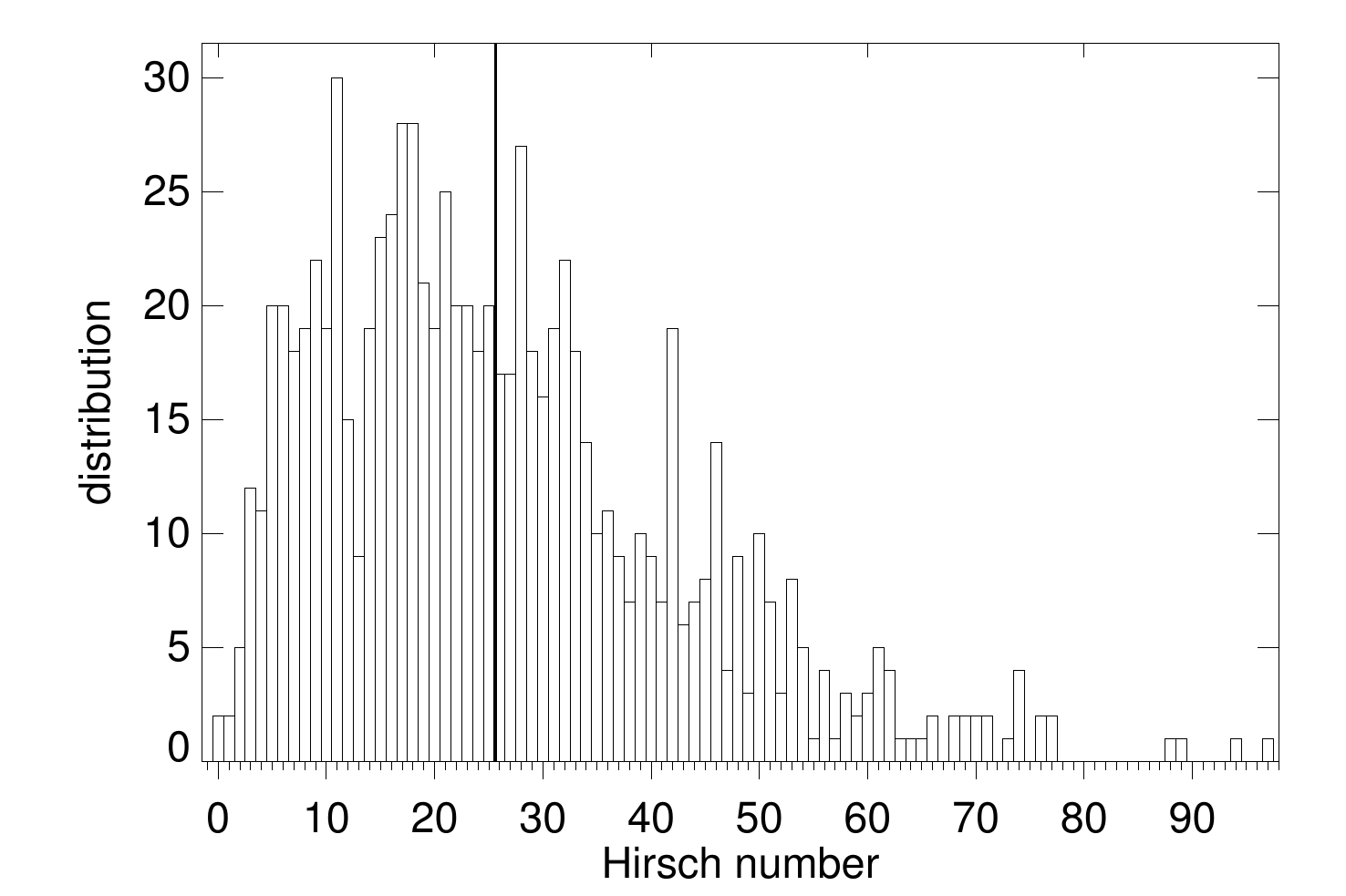} \hspace{-5mm}
  \includegraphics[width=0.5\textwidth]{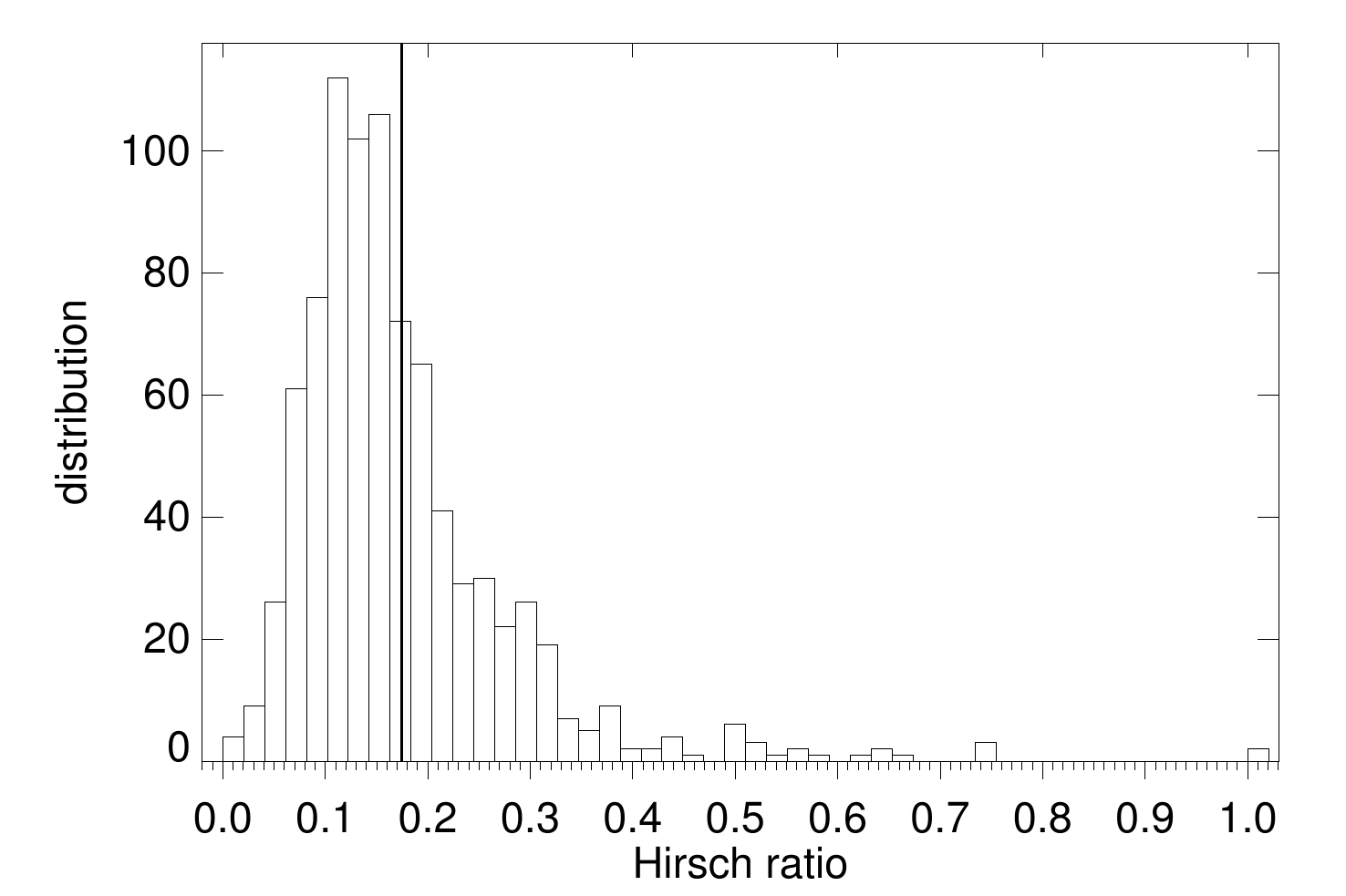}\\
  \includegraphics[width=0.5\textwidth]{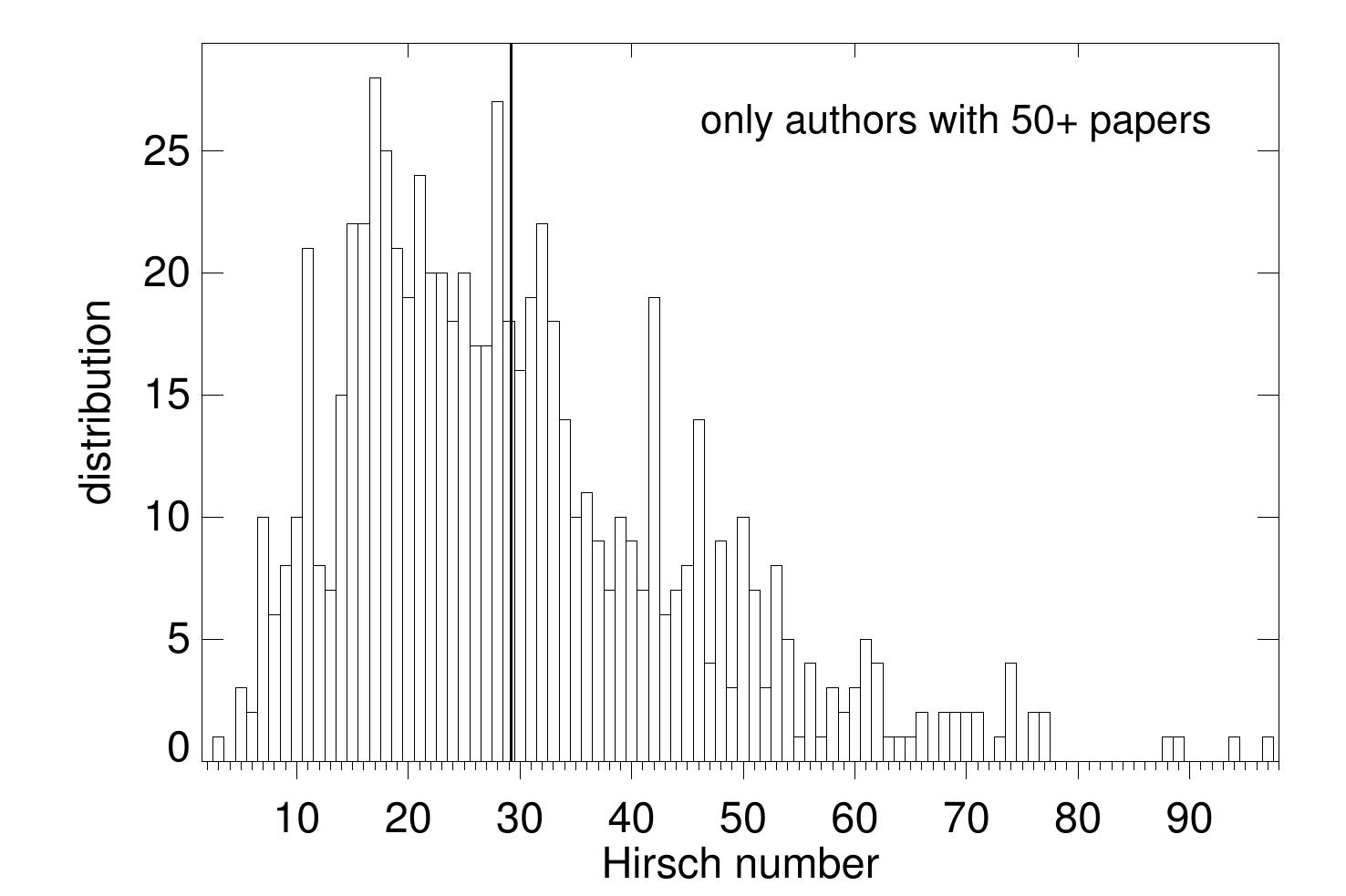} \hspace{-5mm}
  \includegraphics[width=0.5\textwidth]{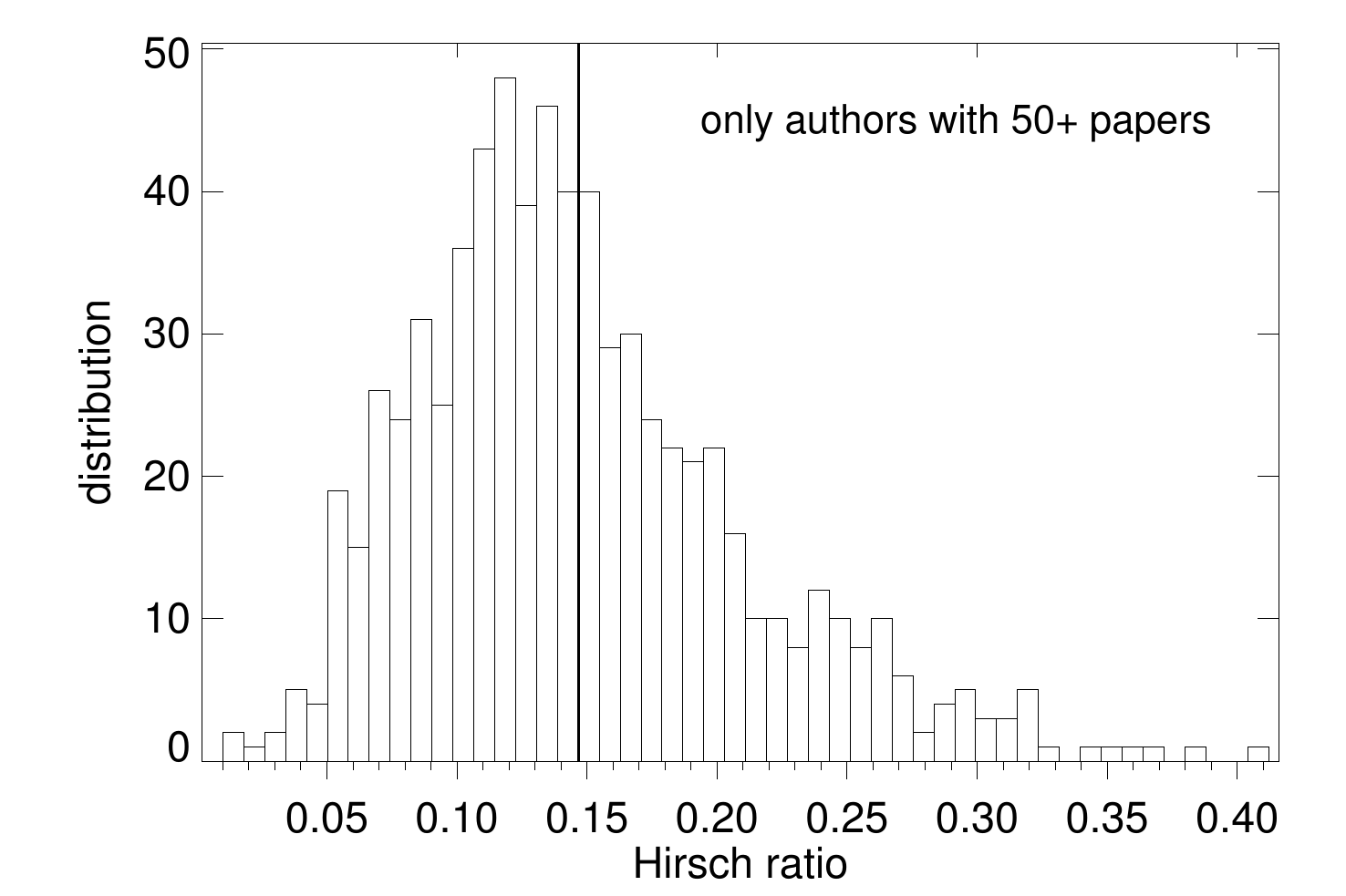}\\
  \caption[]{\label{fig:histograms} 
  ADS-based Hirsch statistics for 800+ solar physicists per January
  2018.
  Upper panels: $h$ and ``quality'' ratio $h/N$ distributions for all.
  Lower panels: only for those with at least 50 publications,
  effectively a limitation to seniors who didn't leave the field and so
  removing the initial hump and bi-modal shape seen top-left.
  Vertical lines: median values.
  }
\end{figure} 

\paragraph*{\dropcap{P}eople.}
Let me start on Sac Peak people of that time by embroidering the
Hirsch theme. 
Table~\ref{tab:hirsch} gives total publication numbers (everything on
ADS), Hirsch indices (the number of publications with at least as many
citations according to ADS), and the ratio of these two numbers.
I collected them for all colleagues that I remember working at or
visiting Sac Peak during our 1977--1978 year there. 
Some had to be estimated due to too many namesakes on ADS (Jack
Thomas, Rich Robinson).
Tim Brown has ADS competition from a physics T.M. Brown but not
producing hirschers galorely like Tim does.
Stirling Colgate's ADS astronomy-only record excludes many
still-secret publications from the H-bomb development (actual bombs, not
Ellerman's 1917 ones)\footnote{He was far too smart for an empty life
as toothpaste millionaire. 
At his Sac Peak seminar: ``I love bombs! 
The bigger the better! 
When I found that solar flares are much bigger bombs than the ones I
was helping to make I moved to them! 
And then I found that supernovae are even bigger bombs!'' At the time
he started a robotic supernova-finder on the Magdalena Ridge beyond
Socorro and later delved into cosmic ray acceleration, but he did not
become a big-banger.}.

Hirsch indices have become the most common way to evaluate publication
standings for scientists; even Google Scholar posts them.  
They grow with time and age, especially academic age. 
They remain low for people leaving the field -- if Einstein had died
in 1906 his Hirsch index would remain three forever. 
They also grow faster nowadays than before ADS made getting and citing
astronomy publications so easy\footnote{ADS magnificently restores ancient
literature to full worth -- as Ellerman's 1917 publication.}.

I add the ratio $h/N$ which I call ``quality'' since it indicates the
mean impact per publication: $h/N \tis 0.10$ (not bad) means one has to
write ten publications to add one hirscher. 
This is a loaded quantity: for Tim Brown a new publication must be cited
twice as often as for me to count -- noblesse oblige! 
Starting scientists publishing their solid PhD publications but no
conference dallies usually reach higher quality than 0.10\footnote{I
began my hirsching when Hardi Peter asked me to give concluding remarks
at the 12th European Solar Physics Meeting at Freiburg in 2008.
I accepted on the condition that I might insult all 200$^+$ colleagues
attending and did so by hirsching all presenters (the pdf still sits
on my website). 
I finished by announcing the two ``best speakers'' based on their
Hirsch: Manfred Sch\"ussler who gave the opening talk, so superb that
I also made him give the opening talk at the Sac Peak Farewell --
where I blundered as his chair when the displayed countdown clock
turned to be out of whack and I hadn't noted when he started; I
cut him 10 minutes short without realizing this even though he reacted
in surprise. 
I did apologize later, getting the gracious response that it kept him
from addressing the ugly solar physics wrongness of the magnetic
Prandtl number (but it sits in his presentation file on the NSO
meeting website which I hope will last longer then the Freiburg ESPM
one which was promised for eternity and which ADS still links for
every talk but has vanished). 
The other best speaker at Freiburg was Tobias Felipe who only had
$h\tis1$ but also $h/N\tis1$, suggesting outstanding quality.}; it
usually goes down with seniority to about this value. 

The high-quality champion of prolific senior types in our field ($N$
above 100) was the late David Hummer ($N \tis 160, h \tis 52$, ratio
0.33), ten times ``better'' than Jay Pasachoff
($N \tis 702, h \tis 22$, ratio 0.03) who instead is a prolific
popularizer. 
The very ``worst'' was Norman Lockyer ($N \tis 302, h \tis 3$, ratio
0.01), discoverer of the chromosphere and helium which he both named
but who lived before the era of citing others ADS-accessibly while
largely filling {\em Nature} which he founded and edited for 50 years. 
So much for calling this ratio a quality indicator!  
The highest ratio in these Sac Peak tables has Pinky Nelson: it surely
takes quality to become a highly successful astronaut\footnote{Pinky
was applying for NASA astronaut selection when we arrived.
He worried about not informing them about his once-dislocated shoulder
-- I guess I can safely spill that secret now.}.

Figure~\ref{fig:histograms} shows Hirsch statistics for 800+
solar-physics colleagues\footnote{I collect their ADS files for years
already, refreshing them regularly in order to have every abstract of
every colleague whose work has ever interested me in my laptop for
easy literature search. 
The collecting engine is a nice script from Alfred de Wijn, with
another collecting Hirsch values.  
The first also collects all the accompanying bibdata specifications
from ADS, so that I never have to produce bibfiles for publications
but instead let bibtex hunt through this entire collection, a nice
trick I highly recommend. 
My website gives detail and the scripts.}.  
Comparison of Table~\ref{tab:hirsch} with these distributions shows
that the Sac Peak staff at that time scores normally (with outlier Tim
Brown taking care of the high-$h$ tail) but that the visitor
distribution was lopsided to high quality, partly by including stellar
types\footnote{Solar physics is a small field so that generally its
Hirsch level is comparatively low. 
Another aspect of this is the relatively wide spread of open solar
data policy, without fierce competition for scarce data between too
many researchers as in \eg\ cosmology, and with Alan Title's wonderful
example for space missions (including his forcing it for Hinode). 
I followed it for the DOT (and also for my teaching and research
materials). 
NSO has open data policy for its NISP monitoring; I hope it will also
have it for DKIST.}. 

Another number that I may add in future is the Hirsch derivative with
time.
Google Scholar sort of does that already by also showing the partial
$h$-index for recent years. 
I expect that for myself it will confirm I am a late-bloomer.

Yet another temporal ratio of interest is ``ache/age'', but for that I
have to find everybody's birth date. 
Reaching unity or larger in this ratio is a tremendous feat. 
I am fairly confident that Tim Brown hasn't reached 76 years yet and
qualifies, but Gene Parker and Jeff Linsky seem to outlive their
hirsch (I certainly do; my ambition is to end up at
$h/\mbox{age} \tis 0.3$).
Other senior colleagues with claim-to-fame $h/\mbox{age}\,\!>\,\!1$ at
present are ubiquitous Sami Solanki, Martin Asplund, Axel Brandenburg. 

I now turn from hirsching Sac Peakers to Sac Peak personalities. 
At that time the staff was still more Airforce (``blue squad'' in
Zirker speak) then NSF; at its start 25 years earlier\footnote{Our
year started with a festive Silver Jubilee meeting "Large-Scale
Motions of the Sun'' including a very posh banquet at the Lodge where
Dick Altrock taught us how to sip a margarita.} Sac Peak was entirely
Airforce\footnote{Early Sac Peak history is described well by
\citetads{1998SoPh..182....1Z}.}.
The joining with Kitt Peak into NSO came later.
Over the years the Airforce contingent dwindled but Steve, bossing
that before becoming NSO boss, kept it artfully alive. 

The outspokenest person in 1977--1978 was Pete Worden, very much
Airforce, always scheming, an excellent host\footnote{His Halloween
party was a screamer. 
He dressed, wigged, and blackened himself into a fat Mississippi lady,
screeching all night in a loud falsetto about her basooms hanging in
the way. 
Some much nicer girls came over from Tucson for the event. 
I got thoroughly drunk but had the Tower next morning and remember
getting there in very bad shape but still in time at sunrise -- to
quickly judge the clouds too thick and go to bed.}, most thoroughly
conservative (``nuke them till they glow'' re Russians after an
Airforce cadre course)\footnote{The summer students had a good prank
while he spent a summer in Cambridge England, posting a letter with
his authentic signature to Zirker that he had now seen the lights of
peaceful liberal democracy and had chosen to stay where those shone so
progressively and socialistically bright, herewith resigning his
Airforce position -- or maybe Jack put it up himself, I don't know. 
The best-ever prank was to paste giant black footprints going up the
Tower -- which Dick Dunn didn't appreciate. 
Later a good one was when Steve Keil returned from a European stay to
find his office, \ie\ its whole content of cabinets, chairs and desk
with all its mess including the stacks of lineprinter output that
adorned any desk at that time, precisely one-to-one arranged outside
in the parking lot with a coke machine and a sofa in the Keil-empty
room.}, a mean volleyball player (but not as bad as Don Neidig whose
serve was a bomb lightspeeding to kill you), and most eager to win any
game in the long winter evenings. 
I had introduced dice poker at Sac Peak, the only game I like (Steve
calls it ``liar's dice''), but the highlight that winter became
Diplomacy, a board game in which you start having a country in divided
Europe in 1914 with an army and/or navy and battle the others in
spring and autumn campaigns, with as distinctive feature that in
between campaigns you huddle in toilets and bedrooms and outside to
make secret deals -- and then betray your newly-accorded partner.
One game took a week of every night scheming and betraying, if not
already negotiated at coffee time in the Mainlab. 
I had expected to be the second slain after Tim Schneeberger, an easy
to slaughter contender in my opinion, but he wasn't and so I was the
very first defeated. 
I also expected that the final battle would be between Pete and
Lawrence Cram and my hope was on Lawrence: also ambitious but more
civilized and likable. 
Not so! 
Surprise!
The final Pete-contestant and eventual winner was Steve Keil. 
I had never expected such tenacious aggression from this quiet
background mumbler, but so I learned that Steve was of eventual
director meat and would be good at that job if that would entail
political win-or-loose fighting, as of course indeed it did and does.

Then-director Jack Zirker impressed me by his directness, efficiency,
quick understanding of anything, and no-nonsense impatience with slow
response. 
But most of all by his American-speak: a natural-sounding fast flow of
idioms I mostly had never heard or read before, always very clear and
very much to the point to make people wrap up and continue to the next
item in our staff meetings.
He came from the Boulder Thomas--Athay eclipse school of NLTE
chromosphere interpretation -- even now an immature subject with
considerable misunderstandings.
In our time at Sac Peak he was mostly administrator while his real
passion was guiding multi-day Sierra Club hikes like the Mount
Rainier circuit.
Later his science was mostly filaments/prominences triggered by
Orrall, with the little mafia of Sara Martin, Oddbjorn Engvold and
pals succeeding in having a meeting in a nice faraway place every
year. 
Since then he put his language skills to write very clear and
well-organized astronomy and physics popularization books: I have them
all thanks to Amazon. 
A good legacy. 

During those staff meetings George Simon was naturally at his most
complaintive, but always pleasant and hospitable to us. 
He made a sport of together walking up the steep little trail from the
Mainlab to the Tower maintaining our conversation, so that I would
arrive out of breath: though short it is very much a thin-air killer
climb. 
Not him thanks to his yoga he claimed, so we followed Pat's one
evening/week yoga class which usually emptied my mind so well that I
couldn't work afterwards. 
Much later I took my revenge on George, together climbing the 14
stairs of the ugly Utrecht campus building where we had moved from the
beautiful but antique Sonnenborgh observatory, now a museum. 
I did that on arrival every work day year-in year-out, and that time
also made sure he talked but not I, and so got him gasping -- but I
didn't boast about my yoga (never since Pat).

By being out in the boondocks Sac Peak had a natural propensity to
collect funny characters. 
Shy types as Ray Smartt and Steve Musman found a good niche there, in
later years Larry November. 
The pivot of good sense was post mistress Marie Cope -- only much
later did I discover Rebecca stems from her. 
She was very nice to us, keeping an eye on our kids gamboling in the
woods behind her office and home; we had given them freedom to range
anywhere within the surrounding roads and they imagined themselves
deep in the Amazon wilderness, building lean-to shelters and becoming
savage bow-and-arrow Indians\footnote{Not hunting lions or bears --
but they might have. 
One summer Rietje was forbidden to take her daily walk along the
airstrip because a young lion was spotted there; I then met a young
bear on my daily bike ride to Apache Point.}.
Marie forwarded tidbits as a thick glossy L.L. Bean mail-order catalog
(``I am sure that Jack Zirker doesn't need two of these and I think
you may like it'') from which we ordered a much-used and beloved
hi-tech Moss tent, air mattresses, and outdoors clothes I still use,
our first experience in shopping-at-a-distance. 
Nowadays Amazon, Abebooks etc.\ serve me well to obtain ancient solar
physics and radiative transfer books not on ADS\footnote{But you might
corner some privately from Andrii Sukhorukov who has scanned the more
important ones. 
I put his scan of John Jefferies' {\em Spectral Line Formation\/}
integrally on ADS with enthusiastic consent from John, but for other
classics getting publisher consent is less trivial.}, right on my
doorstep to flesh out my private library, usually for just a few
dollars plus shipping\footnote{I am so building one of the better
solar physics libraries now that most universities are discarding
theirs. 
My intention is that it eventually ends up in the postdoc
room at Oslo, also a former library without any books left on the
premises but with suited empty bookshelves.  
No point in bequeathing them to Utrecht University which scrapped
its ApJ and A\&A subscriptions the moment Utrecht astronomy was
killed.}.
I don't know how well the relatively complete Sac Peak library has
fared in its move to Boulder; I hope it remains as accessible as it
was, the best I have perused. 
In our year Nancy Carson was the librarian, successfully wooed by Pete
Worden (I was surprised at first until I realized that she managed
well to prick through his boisterous bolster -- ``she will make a good
general's wife'' was the communal opinion). 
The other wooing was of Jack Zirker's secretary Christy Ott by Tuck
Stebbins, also tenacious and also successful. 
I also suspect that Dick Altrock was already wooing Sally away from
Don Neidig (or reversely), also successful.
Sac Peak had been infamous for musical chairs but the couples present
or paired that year were mostly stable.  

Tuck Stebbins and Tim Brown were disciples of Henry Hill who was a
coworker of Robert Dicke at Princeton before he set off to build the
SCLERA telescope on the Catalina mountains near Tucson to confirm or
disprove Dicke's scalar-tensor alternative to general relativity. 
I knew all about that, having dabbled in pole-equator center-limb
temperature differences using the \MgI~4571 LTE line with the Utrecht
spectrograph, not knowing that the Dicks (Canfield and Altrock) had
already done that at the Big Dome; the intention was also to prove or
disprove Dicke's oblateness findings at Princeton. 
Hill's telescope was meant to measure stellar light bending outside
eclipse but I don't think it ever did; instead he did refute the
oblateness\footnote{I believe that the oblateness controversy killed
Dicke's much-deserved sharing of the cosmic background microwave
radiation Nobel Prize (best ever); it was he who told Penzias and
Wilson what they were about.} and detected long-period limb waves that
became a saga of controversy on their own\footnote{Like most low-$l$
helioseismology. 
SOHO was put in L1 for this but also without success. 
When I grow cynic I may write a scathing history.}.
His telescope design was very clever and included Schupmann
correctors; when G\"oran Scharmer put a Schupmann in the SST (after
copying Dick Dunn's turret for his preceding SVST) I was about the
only one recognizing the design.

A result of following these Dicke affairs was that I had learned to
recognize former Dicke pupils as sharing his mold: types who combine
cleverness bordering on genius with utter obstinacy in wanting to
prove something truly outrageous with unheard-of techniques. 
They include Jim Brault, Jeff Kuhn, Phil Goode, Ken Libbrecht, and
effectively but mellower Tuck and Tim as secondary offspring via Henry
Hill. 

Jim's FTS at the McMath was genial, although the solar physics appetite
for $10^6$ spectral resolution without spatial resolution declined
after Stenflo's fluxtube and second-spectrum interests produced its
best results and Livingston, Kurucz and Neckel produced much-used
spectrum atlases.
Jeff (Sac Peak summer student in 1978) is going strong, now on to
exo-civilizations.  
Phil got the world's largest solar telescope named after himself, even
without first becoming a dead senator.
Ken revamped low-$l$ helioseismology but then departed to greener
pastures of more esoteric physics. 
Tuck ended up in LIGO and LISA, humanity's most daring enterprises. 
Tim designed his ``Fourier tachometer'' during our year at Sac Peak
with help from Jacques Beckers who ran tests on a wide-field Michelson
interferometer in his Tower office (the largest there -- in the
Mainlab he had the smallest).
Little did I realize how clever the idea was and how far it would go,
nor how far Tim himself would go. 

Tuck and Tim were both very nice and helpful\footnote{Tuck even taught
me how to read octal memory dumps from fatal aborts (computerese of
that time), a language I forgot since as completely as classical Greek
learned at school.} colleagues, Tim the more reserved but also the
more agile volleyball player, a most important asset in that
volleyball-dominated society. 
Tuck made a winter sport of cross-country skiing the already ancient
former logging-railroad routes, nicely well-graded and yet without
trees growing in the way; I wish I had followed his example and
learned how to ski -- probably the best way of enjoying the marvelous
grades in their then still pristine state. 
Much later Steve and Pamela and me mountain-biked a bunch of them from
the 64 to Bluff Springs but too rocky and bumpy; in the meantime a
number has been hopelessly spoiled by quads following on ``rails to
trails'' motorbikes. 
More on these grades below.

\begin{figure}[hbtp]
  \centering
  \includegraphics[width=0.8\textwidth]{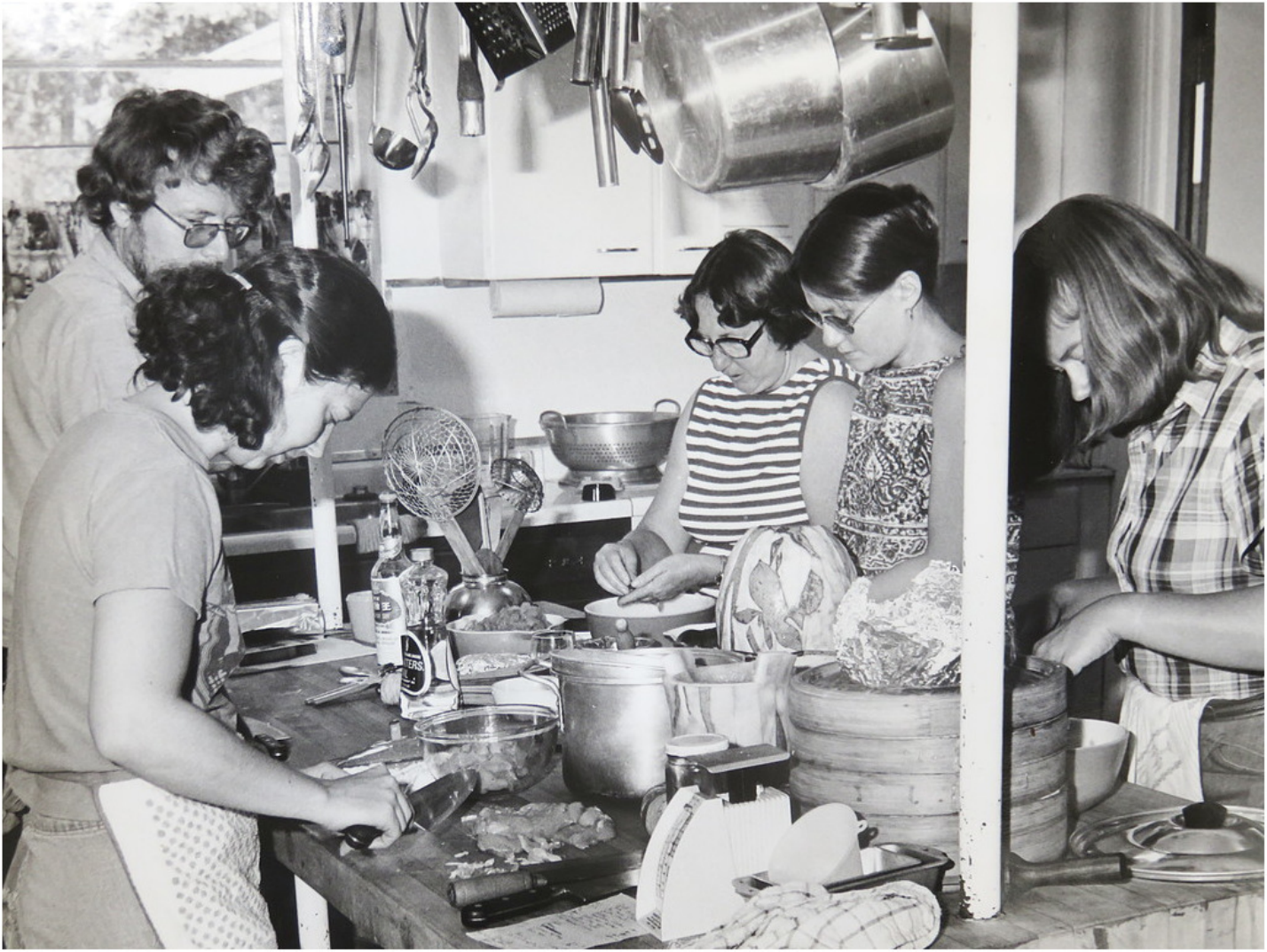}
  \caption[]{\label{fig:chinesefood} 
  Chinese cooking class led by photographer Dick Faller.
  Left to right: Steve Keil, Betty Derrig, Lorette Zirker, Christy
  Ott, Rietje Rutten. 
  This photo appeared in the Alamogordo Daily News for which Lorette
  was Sunspot reporter and was taken during two days preparation for
  the sumptuous exam supper (jellied lamb\,--\,tea leaf
  eggs\,--\,asparagus salad\,--\,radish salad\,--\,Szechwan bean
  curd\,--\,sweet and sour cabbage\,--\,soy sauce chicken\,--\,siu
  may\,--\,tangerine chicken\,--\,roast pork strips\,--\,Szechwan
  dumplings\,--\,sherbet\,--\,stir-fried beef and snow
  peas\,--\,steamed chicken and Chinese sausages\,--\,hot and sour
  soup\,--\,fruit) for spouses as examiners who happily rotund passed
  them all.
  Dick Faller's other passion was classical music, including driving
  to Santa F\'e for string quartet concerts, six hours each way and losing at
  least one car in deer collision.}
\end{figure} 

Franz-Ludwig Deubner was spending a year at Sac Peak when we came and
returned to Freiburg around Christmas. 
In the meantime we made music. 
I had played flute during high school but discarded it at university;
now I had time on my hands and took it up again, buying an old school
flute from Karin Mauter and repairing it in the workshop. 
At Deubner's farewell party at the Zirkers he played one of Bach's
cello suites; I played Debussy's Syrinx. 
Before Franz had been playing all six suites throughout an overcast
day in the Tower whilst every five minutes monitoring the table
rotation precisely in order to find its pivot eccentricity in the
``divided circle'' calibration that both he and I had been taught as a
key effort of classical astrometry.
When he and family returned to Freiburg we bought his piano; when we
left we sold it to Harry Jones; he still has it. 
Much later I made Franz give a serious concert with pianist Pierre
Mein (both had hesitated between conservatory and astronomy), with as
highlight the Debussy cello sonata as a special request of Kees Zwaan
in whose honor the concert was held (1993 Soesterberg ``Swansong''
workshop).  
Franz left Sac Peak with the understanding that he might return soon
as staff member, but instead he became professor at W\"urzburg and
stayed there.
He died last October, leaving strong and good memories.

Lawrence Cram was also already there when we came, with Barbara and
infant Edward.
He was a friend already; shortly before we had hiked the sea floor at
low tide from the North Sea coast to an island (German in fact) and
back at midnight during the next low tide; the roughest, toughest and
perhaps riskiest hike one can do in the low countries.
My boasting about these later motivated Bruce Lites to try one too,
but then it was already November and although we dressed in wetsuits
borrowed from my kayak pals it was bitterly cold to slosh many hours
through winter-cold sea water in the mandatory hurry before it would
rise over our heads again.

Steve and Alice Keil arrived around Christmas from Australia; they had
lived at Sac Peak before and maybe they already had in mind never to
leave again.
They quickly became good friends and they quickly established
themselves as social core at Sunspot -- which they also remained ever since.

A very important Sac Peak asset was the large flux of visitors, not
only from HAO to develop or observe with David Elmore's Stokes-II at
the Big Dome but effectively from all over the US.
Probably there were more than I now remember and collected in
Table~\ref{tab:hirsch}, but even this subset illustrates their variety
and quality.
Sac Peak was an isolated observatory out in the boondocks but a busy
crossroad of solar physics. 
Much later I had similar experience at the SST on La Palma where I
spent five summers educating Utrecht and European students and met
many more solar colleagues than at Utrecht, even running near-daily
seminars featuring SST observers\footnote{This style of tutorial
teaching was the best I have experienced. 
The students would come in pairs and earn five study points by
spending a week beforehand to study their topic, two weeks on La
Palma starting with them giving a seminar on what they had read, then
researching their topic using DOT data, and finally after their return
two weeks more to wrap that up and write their report. 
We sat as a trio at the antique table in the SST library with a
whiteboard and projector at hand; at questions as ``what are
granules?''
or ``what are $p$-modes'' I would give an impromptu lecture from my
ever-growing collection of pertinent display material.
Some of them remain in solar physics, others are now nighttime but
with a high opinion of solar physics, and all those displays still sit
in my laptop. 
It is a shame that the program stopped when Utrecht University stopped
its financing together with the DOT financing after first retiring me. 
In the US it is forbidden to fire people for their age, making me
jealous -- but of course it became worse when Utrecht subsequently killed
all its astronomy, and with it effectively all solar physics in The
Netherlands except CME interest per LOFAR.  Utrecht University's only
remaining astronomy activity is serving my website, still a
champion in download volume.}. 

A clever organizational trick was that for every visitor Christy would
post a sign-up sheet on the main billboard in the Mainlab corridor,
listing all lunches and dinners during the visit with the intention
that one should enter one's name to host the visitor and that he (I
don't remember a she) should be booked completely. 
An excellent mechanism.  
It made me invite people I wouldn't have dared to invite otherwise --
Grant Athay and wife, Gene Parker and wife, Dick Thomas. 
Rietje was nervous about the fearsome latter\footnote{An essay
``Epsilon'' on fearsome Dick for a memorial book edited by his widow
sits under books on my website.} but Lawrence and Barbara gave
the recipe (``a coke and a steak and he is done'') and brought the
essential desert, apple pie with ice cream.

Some Dutch visitors came for me. 
First Theo and Hanne van Grunsven from HAO where Theo worked on a PhD
thesis with Peter Hoyng\footnote{It didn't work out then, but he
restarted after his retirement and is now getting close, also being
close by living in the village across the river from ours, the same
street as Christoph Keller with his family -- Holland is small.}.
Then at Christmas\footnote{I cut our Christmas tree along the Cow
Trail and spotted the stump years after -- but less intrusive then
Micky Mauter's tree gathering who at Traute's pointing to a
nice-looking high-up tree top would hear, when he had chain-saw-felled
the whole tree and held its top up for inspection: ``No, that isn't
what I thought; maybe that one up there is better?''} Henny Lamers with
family from JILA and Henk Spruit from HAO for our joint vacationing in
Mexico. 
My mother came in spring. 
The most unexpected was Aloysio Janner, professor of theoretical
physics at Nijmegen University, on what one now calls head hunting to
get me to Nijmegen to revive astronomy there. 
He was first vomited upon by Martijn\footnote{Martijn necessarily
shared Rietje's every-day snow-chained drive to Cloudcroft to pick up
Katrien after kindergarten and daily vomited in the many curves of the
Sac Peak road.
Katrien took the school bus in the morning with the Altrock boys but
it returned too late; Karin Mauter already drove her own car to
school. 
The road was the second one, not the original one up through Water
Canyon that is now favored by long-stay campers in August and was then
favored by Jack Zirker for painting landscapes -- many SacPeakers
painted the forest, Don Neidig and Alice Keil the best (an Alice aspen
pair bought for our 25-year marriage adorns our dining room); Christy
Ott was into art photography (a canoe triptych adorns my Lingezicht
study -- known as ``the room with the books'' by the villagers);
earlier there were famous Sunspot pottery classes. 
Later Scenic Byway 6563 (numbered by George Simon) was mostly laid over
our road, doing away with the dangerous cattle guards in bends
(including infamous Cathy's for Cathy Gilliam) and the difficult steep
Horseshoe Bend. 
I wonder how many cars suffered animal collisions there; next to Dick
Faller also Louis Strous, another Kees Zwaan pupil from Utrecht who
was at Big Bear and Lockheed before Sac Peak but then left solar
physics to return home and find a wife (he did), not having found one
in the Cloudcroft Light Opera Company -- maybe because he always acted
the villain. 
He also started an ``Ask Mr. Sunspot'' site that he continues as
``Astronomy Answers'' (\url{http://aa.quae.nl}).}
and reacted graciously with a welcome package of Lego\footnote{Our
children were lonely at Sac Peak. 
During the winter there were no others of their age but that improved
by summer.
Katrien quickly learned English (New Mexico style American) at
kindergarten and taught Martijn; in spring it became their primary
language. 
At first they used Dutch idioms in English (``No, I don't can play
now''), then reversely (``O, ik zie!'').
We also learned from Katrien: after a week she asked what a PB\&J is
that she got offered in breaks -- we had to ask the neighbors.  
She later assisted me in guiding Saturday observatory tours for
Texans: ``My daddy knows more difficult words than I do, but he has a
very bad accent and I will help explain''. 
It is a pity that we returned home just before she learned to read or
her skill might have persisted; instead, they quit speaking English
together immediately because the other children in our village school
made fun of them. 
In the meantime Katrien's English again became far better than mine,
but King's English now.}.
He was much impressed by the Tower telescope, by the research being
done there, and apparently sufficiently by me to offer the
professorship. 
However, when I negotiated at Nijmegen after my return I declined when
it became clear that I would only be permitted bachelors students, not
masters students -- the aim was to attract students initially by
offering sexy astronomy but not at all to distract them from serious
physics eventually. 
Decades later Jan Kuijpers, colleague solar physicist at Utrecht,
became professor there with wider empowerment and did revive astronomy
there, expanding it into a highly successful institute that after
Utrecht's demise also accepted most of the Utrecht staff -- except
ex-NSO Christoph Keller whom I had snared to come to Utrecht as my
effective successor, was director at the time of the Utrecht
closure\footnote{See his account of this sad affair in
\citetads{2013ASPC..470....3K}.},
and took his group then to Leiden and into exoplanets, and Alexander
V\"ogler from G\"ottingen, my nominal successor appointed by
Christoph, who became a piano teacher.

\paragraph*{\dropcap{H}ikes.}
Weekend hikes became an important aspect of our life in Sunspot and
a most valued asset of the location. 
Naturally Dog Canyon became the champion -- by now ten times. 
The first and last were disasters.

The first was soon after we arrived, having Henk Spruit visiting from
HAO. 
Pinky Nelson asked Steve Musman to guide us. 
We started at the Tower and scrambled straight down the slope avoiding
nasty cactus to the West Side Road and got on to the Joplin Ridge
Road.
When we questioned Steve where we should go left into the canyon that
we saw enticingly from above he didn't know, so we again scrambled
down the slope with Henk reconnoitering cliffs blocking our way into
the canyon and spotting detours around them. 
There we ate our lunch and drank our water in the shade of the wall,
believing we were already at desert level just as Pinky and I had been
on an earlier hike down Alamo Canyon with Jack Thomas and family.
However, Steve then started muttering about a waterfall. 
When we got to the top of it I saw our car glinting with my
binoculars, far away and far below us, but we saw no way down and then
spent much time scrambling around the ledges to find one, loosing each
other with Steve whistling most confusingly because he wasn't where
his echos sounded. 
We finally bunched up again and went seeking the Eyebrow Trail by
going up a northward side gully and west over the adjacent hills. 
At the second one, just before we would indeed have seen the trail
below us, Pinky decided decisively we must stop searching but take the
fastest way back up -- our water was gone, it was hot, and it was
getting late.
So we went up the slope, retrieved the Joplin Ridge Road, made the
West Side Road at dusk, and then Steve remembered the existence of the
Cow Trail. 
I didn't believe him but he found it and it proved a godsend in the
dark, with a roll of Live Saver candies in Pinky's pocket earning
their name.
We finally stumbled home near midnight, finding Franz Deubner
babysitting replacing his daughter whom we had promised to be back at
four o'clock.

The next time we checked those hills above the waterfall, establishing
what Steve Keil called the ``Rotten Rutten Route'' ever since -- on
his yearly ``death march'' he preferred to rappel summer students down
the waterfall.
We also followed my route at his 1983 workshop because he was worried
about Nigel Weiss who himself was more worried about his son. 
In the upper canyon we met Bruce Lites and Larry Wilkins going up,
having chimneyed up the waterfall after depositing cars at Oliver Lee
for our return.
I then realized that going up is a much better idea than going down;
the next seven times I led the hike doing this along my roundabout but
very scenic route: up the Eyebrow Trail, cut right well above Indian
Cave and scramble over two wide dry hills after first climbing a
scalable ledge, descend into the east-bending side gully seen far
below, follow that until its junction marked by house-sized rock
pillars, and then at the mouth besides the waterfall follow a narrow
ledge and then scramble down into the upper canyon upstream of the
waterfall. 
One time going up with Oskar Steiner there was a strong cat small in
the clump of tall trees there, but we didn't see a lion. 
Then follow the long upper canyon upstream. 
Twice its stream was actually running and swimmable in pools (I won't
insert my naked-colleagues photo here), the second time even full of
mature fish which was puzzling because the creek runs only rarely --
until we found that the dam of the new fish pond at the top had been
flooded.
It belonged to the isolated houses built there in the 1990s that Steve
warned must be skirted wide away from the ferocious dogs. 
The last time there I pulled out the Alamogordo Daily News of that morning
with headline ''Killing at Dog Canyon'' which rather upset my
companions until Alfred de Wijn read the article: ``Oh, just a man
shooting his wife in the housing below that we drove along, that's
perfectly normal, no worry, we are OK''.
As before we then hid ourselves along the dip of the southern
creek until we got tree cover to cross over and regain the northern
creek. 
Then follow that up along its many curves, ignore two side roads to the left,
continue all the way up to the West Side Road, from there (unposted
but its entrance marked by anti-bike rocks) finally up the
ever-steepening Cow Trail -- a suited companion contest or killer.

In doing these trips I got into the guide's habit of inspecting my
group at the start and wonder who was going to be the problem. 
Always another one than you think; certainly the last time because
then it was me. 
I barely made it -- I lost my step in Dog Canyon (not Nantucket), indeed
on the Cow Trail where Andrej Tlatov and Maria Loukitcheva
patiently pushed me on to eventually reach the top at dark.
Thereafter I could barely climb our home stairs, but presently having
two new bionic knees I hope to hike again -- although Dog Canyon
remains beyond my bucket list.

My largest other hikes during the reminiscence year were from Phantom
Ranch up to the South Rim with Henk Spruit after rafting down-river from
Page (we couldn't pay for more then, but since my retirement I have
done the whole stretch to Diamond Creek four times, duckying all
permitted rapids including Hermit), the Barrancas del Cobre in
Chihuahua (of similar size as the Grand Canyon but less well
known) with Henk and Henny Lamers in our midwinter vacation, and in
summer from Three Rivers all the way up Sierra Blanca with Steve Keil
and Jeff Kuhn through beautiful unlogged forest, spending the night
under the first trees down from the top in my coldest night ever.

\begin{figure}[hbtp]
  \centering
  \includegraphics[width=0.8\textwidth]{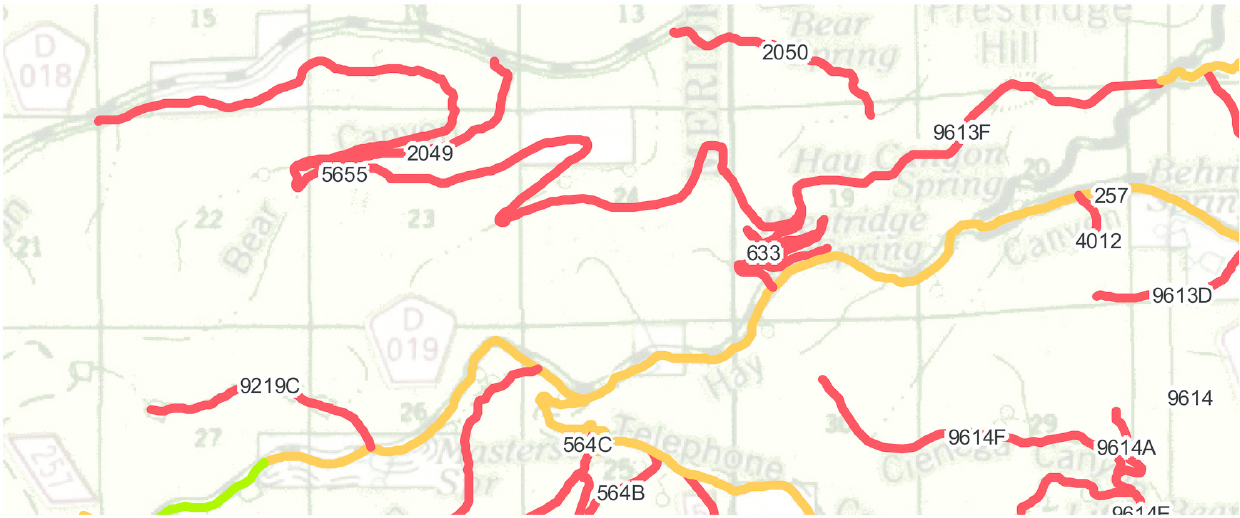}
  \caption[]{\label{fig:switchbacks} 
  Logging railroad switchbacks. 
  Extract of a National Forest map showing hiking trails following
  ancient (1930s) logging railroad grades in red. 
  Long before these trails got numbered I noted the five switchbacks
  of 633 descending to Hay Canyon on the western slope of Prestridge
  Hill on an aerial photograph in an excellent railroad history
  brochure found at a Cloudcroft fair (Fig.~24 in Vernon J. Glover,
  ``Logging railroads of the Lincoln National Forest'', out of print
  but there is a downloadable scan$^*$).
  We retraced it in 1993 on a long hike with Phil Wiborg, down to
  Wills Canyon from the 64 and then going up along what is now Trail
  5655. 
  I got worried when Phil said in Bear Canyon that the big rocks on
  the trail that looked like have been overturned recently were a sure
  sign of bear looking for ants, probably just ahead of us. 
  Later we went up the switchbacks (their bottom is or was marked by
  the remains of a watering structure at the creek on the other side
  of the road) as family outing, first with Phil's newly acquired one,
  then the Uitenbroeks with Kit Richards and later the Reardons,
  descending along the grade not shown here that branches off at the
  top (at the red U turn) and goes down to Hay Canyon further west.  
  The last time was with Han Uitenbroek and Ali Tritschler, again
  coming up 5655 -- not seeing bear sign in Bear Canyon but an eaten
  elk in the next U turn, lion? -- and then continuing along 9613F
  (demolished by quads) to the Prestige Hill Road where we had hidden
  the beer the three of us enjoyed while Han ran back along the roads
  to regain his car. 
  Each time it felt like an archaeology expedition -- although back in
  history only to the year I was born; soon after all rail was taken
  up to become Liberty ships to free Europe including me. 
  It is all gone, one only finds a few leftover bolts and ties,
  plus some marvelous trestles falling apart. \\
  ($^*$) {\scriptsize \url{
  https://foresthistory.org/wp-content/uploads/2017/01/Logging-Railroads-of-the-Lincoln-National-Forest.pdf}}.
  }
\end{figure} 

In later years our most boring hike was the Rim Trail from Cloudcroft
to Sac Peak with Nick Hoekzema\footnote{At the 1998 workshop Nick, my
nonconformest pupil, brought his bike and pedaled from El Paso to Sac
Peak, ``rather boring at first''.
Going back he went down the hill from Cloudcroft before our car
(Sebastian Steffens driving Oskar Steiner and me) at exactly the speed
limit, to the Space Hall of Fame where he knew every little detail
about any Apollo astronaut. 
Later he went to Lindau and Mars studies.}, endlessly up and down dark
spruce forest along the trailbike-spoiled trail with almost nowhere a
view.
Amazingly, Matt Penn used to run its full length -- speeding may make
it less boring.

It was much better to discover old railroad grades including the five
impressive hidden switchbacks off Hay Canyon with Phil Wiborg, and
beautiful San Andres Canyon, the northern fork of the two converging
valleys seen below the Tower lookout (with a recent rock slide that
Kit Richards dated per Google Earth), but where we always turned
around at the waterfall. 
There is a marvelous meadow with beautiful tall ponderosas at the
confluence of the two creeks. 
We camped there three times, instructed by Larry Wilkins how to make a
sizable swimming pool with the plastic sheet he hid there for the
purpose by letting it fill a wide rock cleft overnight. 
He and his boys had a good habit of cleaning the trail with chainsaws,
but now there are many newly fallen trees over it and it became harder
to find.  
Larry knew the Sac Peak mountains better than anybody except
possibly Micky Mauter.
Larry once descended the San Andres waterfall by climbing down a
cliff-leaning tree and scrambled around the promontories along the
knees and benches halfway up the slopes, always unexpectedly petering
out, all the way to Dog Canyon, but taking days longer than he had
told Tina and surviving thanks to watering at Frenchies Cabin -- not
an example to follow, but I did put San Andres on the agenda for the
day after the Farewell Workshop and collected a dozen colleagues
daring to get guided by me. 
Not so, they had to find the waterfall by themselves\footnote{It has a
nice bathtub-size pool-with-a-view at its very lip from a choke block
in the cleft, but my friends found it too muddy from the heavy
rains the night before that had also made the West Side road
slippery but luckily didn't repeat.} since I gave up sooner
than I had hoped, returning already shortly after the cabin ruin in
fear of my not-yet bionic other knee, maybe symbolically fitting the
concept of an ultimate farewell hike.

\paragraph*{\dropcap{L}incoln forest.}
Sac Peak is amidst the Lincoln Forest, its major environment definer.
From the desert below this forest appears quite unexpected. 
The Chihuahuan desert is a high dry plain, a former sea floor and
therefore as flat as Holland. 
It was covered by grass prairies around 1900 but those were quickly
eaten by cattle barons only slightly hampered by Apaches chasing up
their cows along the Eyebrow and Cow Trails to hide them in the forest
above and initially successful in stoning the US cavalry coming up
Eyebrow from the cliff above -- until General Custer (a General Pete
type) put an end to their capers. 
Now the Mescalero's do well again thanks to having snow for Texans to
ski on Sierra Blanca in winter and casino rights for milking Texans
in summer. 
In the meantime the desert got covered by thorny scrubs (creosote bush
and mesquite) and became unattractive, as boringly evident 
on the long drive up from El Paso where the only pastime is to spot
antelopes (pronghorns), be the first to spot the Sac Peak tower and
nowadays also the telescopes on Apache Point, and to recognize the
narrow indistinctive entrance to crooked Dog Canyon behind Oliver
Lee visitor center. 
The only Tularosa basin features of interest now are uniquely splendid
White Sands, the adjacent lava flows best seen near Carrizozo, the
petroglyphs at Three Rivers, occasionally rockets launched over
White Sands, and weird aircraft stealthily climbing from
Holloman.

The secret of the Lincoln forest is watering from catching snow in
winter and thunderstorms in August even though most vanishes into the
limestone, possibly into undiscovered caverns below. 
The main runoff is the Penasco River but it is mostly a meager creek,
only way down east large enough for Steve to fish and our kids to
float, activities we tried combining but they didn't match.
Flowers do not come out before summer, spring means just snow
evaporation to expose brown muddy earth, a large disappointment
to Europeans eagerly expecting a springtime flower carpet. 
The desert below did better in cactus, especially in the 
Sonora desert when we drove to Tucson in March.

In Alamogordo the mountains pose a forbidding barrier to the east, but
one doesn't appreciate that their tops appear black rather than
limestone-white because they are densely forested with ``Canadian
zone'' vegetation. 
East from the cliffs of the Tularosa fault they descend slowly towards
Texas as rolling pine-forested hills that indeed look like Canada. 
New Mexico mountain tops are ``sky islands'' of cooler and wetter
climate; the Lincoln Forest on the Sacramento Mountains is a prime
example.
The Guadeloupes seen from Apache Point to the southeast, hard to reach
from Sac Peak but very worthwhile, are less high so drier but more
remote.  

Starting to drive up the 82 to the crossing with North Florida Avenue
(with its last gasoline station where the very first evening the
cashier girl, my second girl encounter in the US, asked: ``You have a
funny accent! 
Are you from Oklahoma?'')
one may first detect (with binoculars) the Sac Peak Tower on the far
rim\footnote{The Tower even sticks out in the view from Sierra Blanca
and also from White Sands and from the Holloman Base sewage lakes off
the I-70 where one may spot actual ibis, not an Italian artifact.
They glint in the evening seen from Sac Peak. 
``The officer's lakes'' Horst Mauter said when asked what those were,
maybe because officers bullshit more, but they are cleaner now and a
birdwatcher paradise at migration times although I think ibis are
there year-round and breed too but not publications.
Take the abrupt binocular-signed turnoff after the last little hill
along the I-70 before White Sands and go right after the first deep
lake, over the levee to the shallower birding lake.
The only location where we saw the Tower sink below the higher
Sacramento hills to the east is recommendable Aguirre Spring
Campground below the Organ Needles on the way to Las Cruces, up the
opposite side of the Tularosa basin. 
We camped there again the night before the Farewell with Roman and
Natasha, our longtime friends from Kiev who had cornered Trump-era
visa thanks to a well-phrased invitation letter from Valentin (one
Iranian and two Chinese did not get in) and whom we showed the
emptiness of New Mexico before the meeting but getting stuck on a dirt
road from the VLA to the Gila Cliff Dwellings.} and also the trees
there that blacken the upper reaches. 
On the steepening way up one then passes through dry yucca and agave
canyons, after the tunnel\footnote{New Mexico's only tunnel became an
organizational crisis for the Farewell workshop when it was belatedly
discovered that it would close every night that week at eight o'clock
for long-overdue repairs. 
The LOC's preference to lodge participants down in Alamogordo rather
than up in Cloudcroft (I contested but it took long before Cloudcroft
options appeared on the website) made them suggest to scrap the
Tuesday barbecue and traditional Sac Peak volleyball game, the
Wednesday White Sands excursion, and that the Thursday banquet should
be down in Alamogordo. Luckily Alexei Pevtsov pointed out, his main
act as LOC member, that there is a a suited alternate road connecting
La Luz to High Rolls. 
I google-mapped it after a helpful hint from Jacques Beckers and saw
that it is paved and even has a center stripe for night driving.
It did suffer a no-pass flash flood shortly before the workshop but
not during it and in the end we did have Tuesday snacks and an
impromptu volleyball game, the Wednesday White Sands excursion, and
the Thursday banquet at the Cloudcroft Lodge (well, at a dismal shed
and Alamogordo fare might have fared better).}
one gets into High Rolls oak scrub and fruit trees, and then one
drives up along the former water faucet where we used to collect our
drinking water, Sac Peak's water from the water tower (never climbed
by me but by most summer students although forbidden) being too
chlorinated, on the right just before the A-frame still there, and
then one ends up in the tall-pine Lincoln forest starting at the
railroad-trestle level.
A most amazing ride up into an uncommon inverted upside-down treeline
(in the Alps one skilifts up to get above trees, here one drives up to
get into them excepting the ski lifts on Sierra Blanca). 
However, the Lincoln forest on top is now largely spoiled, irrevocably
so.

We never realized how much until Tower observer and avid naturalist
Mike Bradford (deceased 2017) made us hike up San Mateo mountain
across the Rio Grande. 
It also reaches high enough to catch forest-growing snows and
thunderstorms, but it was never clean-cut thanks to being small and
steep. 
On that climb, indeed steep, one sees what the Lincoln forest must
have looked like before the railroads enabled total-cut logging:
magnificent open forest with giant but well-apart ponderosas
interspersed by neat and tidy clumps of tall pine and aspen, with
saplings everywhere but mostly to be burned in the occasional
natural wildfires that remove undergrowth while tall trees survive. 

Instead, the clearcut harvesting of most of the former Lincoln forest
has now translated into very high danger: the overly dense (orders of
magnitude overdense) mono-culture same-age impenetrable extended
forests of black spruce one particularly encounters along the former
railroad grades (not at Sac Peak itself where the trees were thinned
for a fake sense of security) predict far too hot forest fires in
which everything burns down completely, causing repeat of this ugly
too bare $>$ too dense $>$ too bare cycle, as can be seen already in a
tell-tale burn stretch along the 64\footnote{The same happens with the
splendid pine forests covering the volcano top of La Palma where the
pines also depend on occasional fires for survival, even for shooting
off their seed cones as jet-driven rockets, but now burn all-out
thanks to the EU furnishing millions of subsidy for fire-fighting
roads, faucets and even sprinklers all over the mountain.}.

\paragraph*{\dropcap{R}eturns.}
Steve proposed that I should write this reminiscence under the title
``Why I keep coming back''. 
Simply because I love the place and so does Rietje. 
Sadly, it will be ``Why we kept coming back'' from now on. 
For Steve and Alice this title would simply be: ``Why we never left''
-- but they did eventually, Pamela even to the far side of the world.

I came back alone for the outstanding workshops\footnote{I think that
Lawrence Cram proposed in the spring of 1978 to hold yearly Sac Peak
workshops in the format of the Silver Jubilee the autumn before.
The NSO website serves the proceedings or links to them from 1980 on. 
That one was Dick Dunn's ``Solar instrumentation: What's Next?'' which
sports Hammmerschlag's open DOT tower on the cover, but not yet on La
Palma and not yet with a DOT on top -- that would take Rob another
seventeen years. 
In the meantime Rob had parked the tower at the Westerbork radio
telescope; Henk Spruit and I climbed it in 1977 on the way back from
our sea floor hike with Lawrence. 
Henk was very skeptic of Rob and DOT and later killed the latter in
very repugnant manner.}
organized by Steve in 1983 and by Bruce Lites in 1984, the latter
followed by a solar-stellar one set up by Stuart Jordan (who joined my
first Dog Canyon up and figures on my not-shown naked-colleagues
photo). 
Then with the family in 1991 after the Mexico eclipse\footnote{My next
eclipse after the 1970 one in the same area on which my thesis was
based; I then started viewing eclipses as tourist. 
By now I have been to 13 with only 1.5 clouded out; the latest 
the glorious USA eclipse after the farewell.}, again per sabbatical with
Rietje in 1993, then alone for the workshops in 1998 (DST naming) and
2000, in 2003 on another sabbatical with Rietje and also Jorrit
Leenaarts as starting graduate student
to learn tricks from Han Uitenbroek, in 2005 alone to the Bob
Steinfest, and after my mandatory retirement (2007) three months with
Rietje in 2008, together to the workshop in 2009, six weeks in late
2013 when at home I became too unhappy being the only active solar
physicist in Holland\footnote{Not counting Daniel M\"uller because
ESA's ESTEC is an international non-tax-paying enclave in Holland.},
and finally the 2017 Farewell: my thirteenth time, presumably the
last. 
Clearly, we love the place -- as it was.

\paragraph*{\dropcap{F}arewell evening.}
My last evening at Sac Peak of the reminiscence year became my
ignominious claim to local fame. 
I had booked the Sigma\,5 to write a magnetic reel tape with all my
programs and data files residing on its disks, very many kilobytes to
take home. 
When walking up to the Mainlab side door in near dark I saw Christy
Ott's black and white cat sitting next to it, as often before. 
As usual I bended over to stroke, but it jumped on its forelegs and
whipped up a huge white tail. 
I realized it wasn't the cat but a skunk! 
I jumped back and quickly went to my office to phone Rietje and boast
how narrowly I had escaped being skunked -- but then I suddenly nearly
suffocated: the fluid sprayed on me was evaporating.
I could hardly breathe, ran outside to gulp fresh air, and tore off my
clothes. 
But I still absolutely had to fill that tape, so I did that in my
underpants and went home in them and buried my clothes in the forest
-- except for the Zeiss binoculars hanging on my belt all that
year\footnote{The first rare US bird I saw through them was a whooping
crane at Bosque del Apache with Theo van Grunsven, to Jack Evans'
dismay because he went there regularly for them but without success. 
Later that year we learned that the Jornada del Muerto drive to the
Bosque can be enriched by turning south at the lonely Bingham rockshop
halfway to camp at the scenic and old-timer-stuff-rich Bingham mine up
the slope where a sizable fraction of the periodic system sits in the
rock pile put up for rock hounds to keep them from venturing into the
shafts and tunnels.
The mine caused a dip in the White Sands perimeter and so is the
closest one can get to Trinity Site outside its visit times.}
and ever since; for years I smelt skunk when birdwatching.

The next morning while I made my farewell round Chuck Bridges asked
who had gotten a skunk into the computer room. 
He had come in after me for his butterflies and concluded from the
stench that there was a skunk underneath one of the many cabinets but
hadn't been able to find it. 
I then made the mistake of confessing it had been me.

Many years later somebody asked: ``Are you from Holland? 
Do you know there once was a Dutch guy here who got skunked right in
the Mainlab?'' -- my Sac Peak fame.

\begin{figure}[hbtp]
  \centering
  \includegraphics[width=0.8\textwidth]{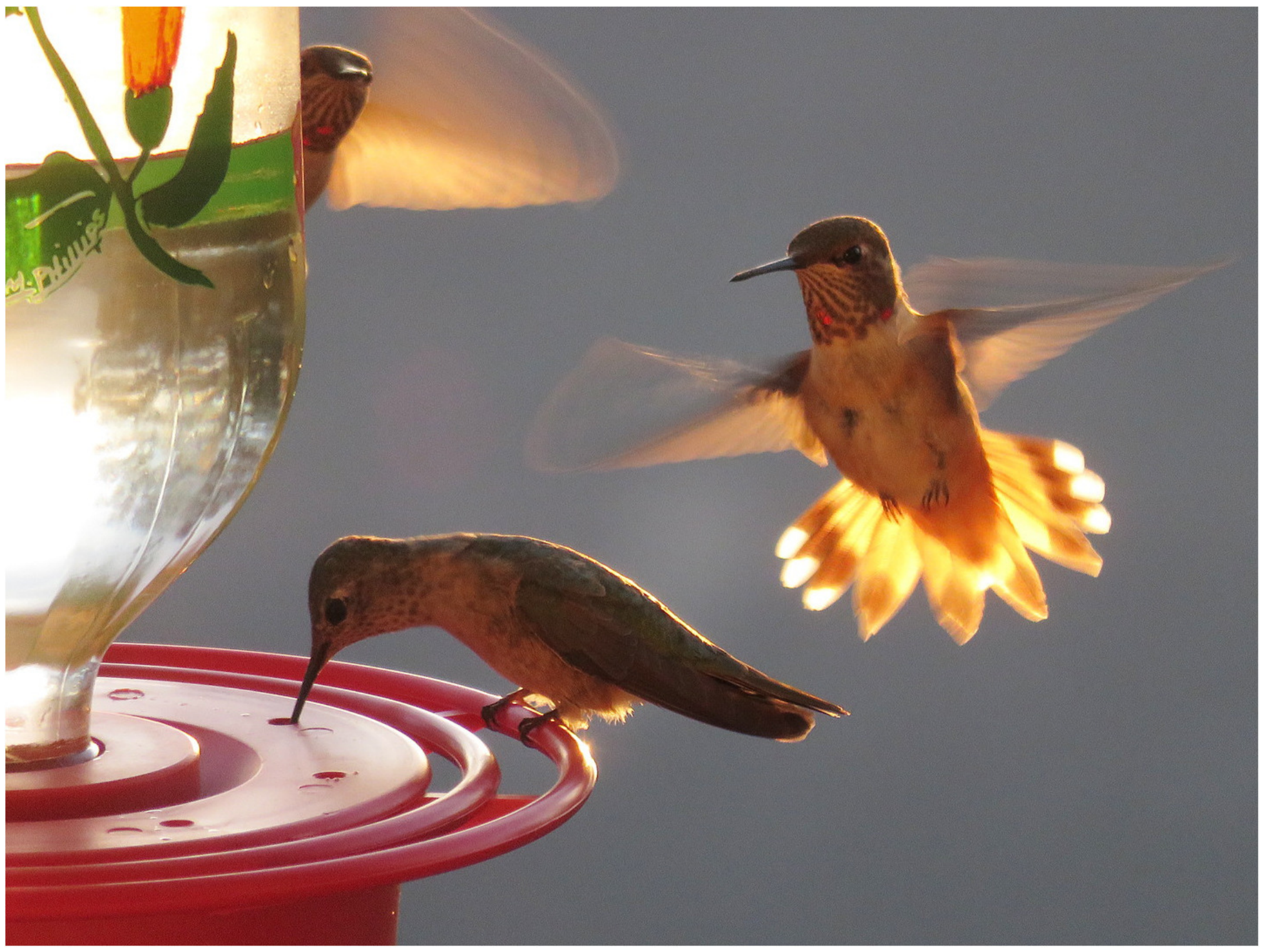}
  \caption[]{\label{fig:hummers} 
  Hummers at Rebecca's feeder outside our VOQ window, Broad-tails I
  think, not Rufous because those are so aggressive there wouldn't be
  another present. 
  Taken after the Tuesday volleyball game that was on the plan
  originally but got scratched -- and then materialized nevertheless
  thanks to Steve having a ball.
  This photograph is nice but I also took videos which are much
  better, although not as spectacular as the high-speed one that I
  still have from Kit Richards using a fast AO camera in which you can
  follow the tumble-over-backwards maneuver by which they roll out of
  the way of a kamikaze-harpooning Rufous. 
  One summer I stepped in for Tower observer Dick Mann so that he
  could vacation elsewhere for a week, accepting to daily boil huge
  amounts of sugar to refill his multiple feeders -- he must have
  multiplied the Sac Peak hummer population at the time.
  Standing between his feeders was putting yourself straight into Star
  Wars: they would suddenly materialize out of some other dimension to
  inspect my nose, then shrilly scream and vanish as instantaneously. 
  I wish we had those critters here in Europe -- whoever missed out on
  them was not an intelligent designer. 
  On Wednesday I didn't join the White Sands excursion but stayed in
  our apartment to practice my own talk the next day (it included
  running IDL live which is risky) but instead I just sat glued to the
  window until dark.  
  The next evening I didn't see any since the Farewell Banquet at the
  Lodge was on the program -- but it wasn't at the Lodge and not at
  all like any of the previous Workshop banquets but a lousy affair,
  perhaps fit for a farewell in the sense of never ever again. 
  I should have stayed home to watch more fulfilling hummers instead. 
  }
\end{figure} 

\begin{figure}[hbtp]
  \centering
  \includegraphics[width=0.8\textwidth]{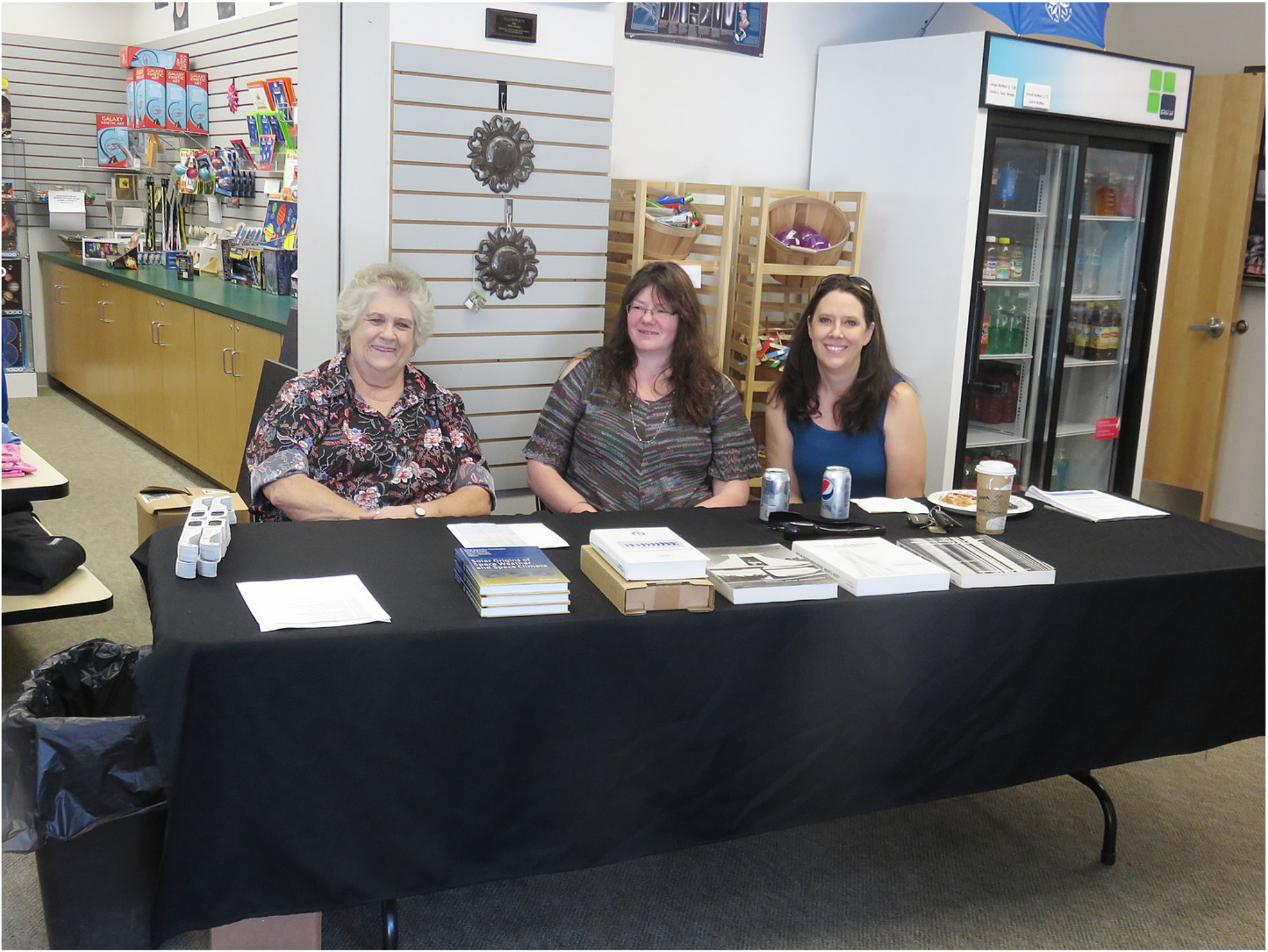}
  \caption[]{\label{fig:workshopladies} 
  The NSO ladies behind the Farewell: Rebecca Coleman, Lou Ann
  Gregory, Jen Ditsler.  Taken on the last day but they were still smiling. 
}
\end{figure} 

\begin{figure}[hbtp]
  \centering
  \includegraphics[width=0.8\textwidth]{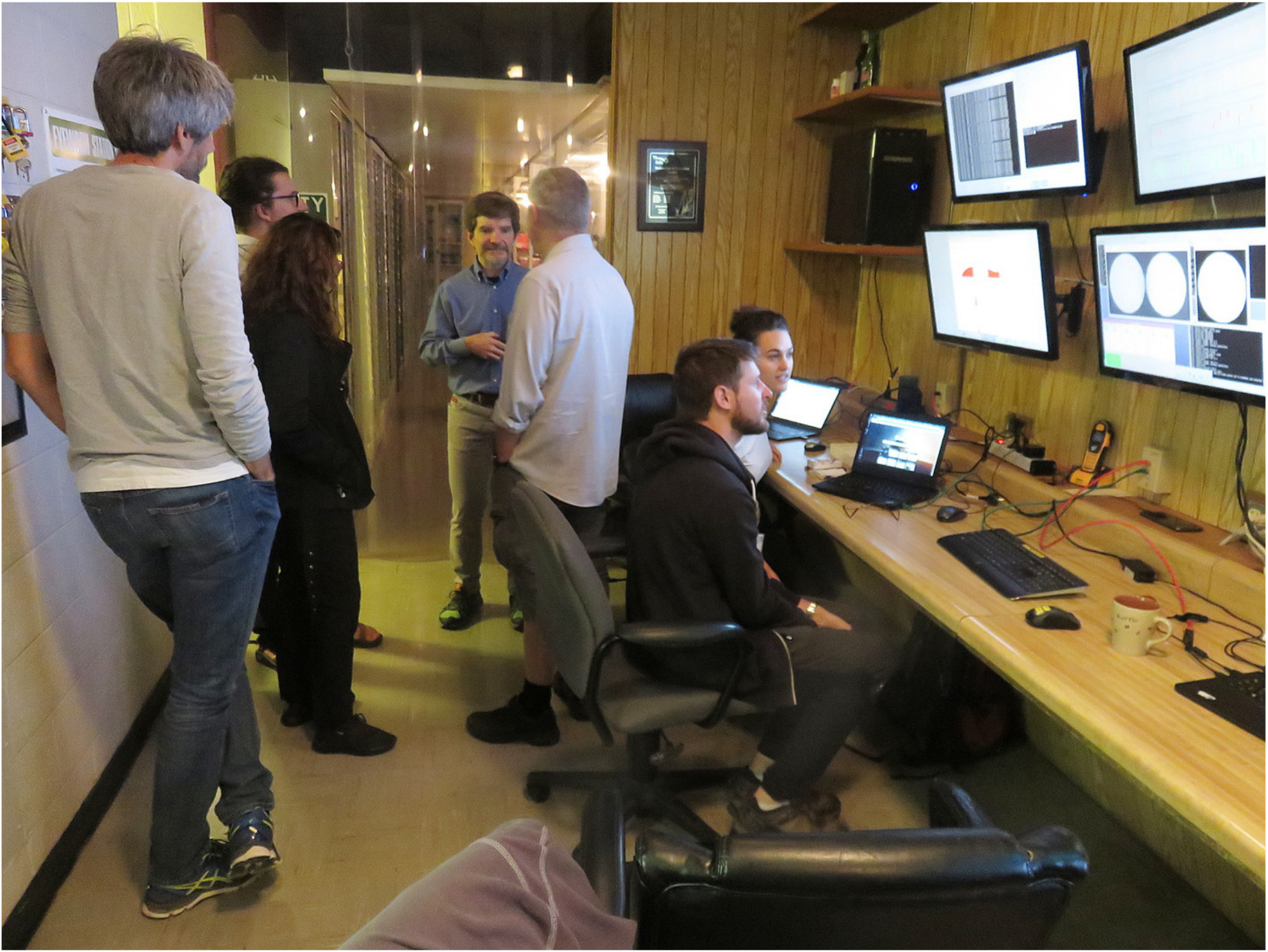}
  \caption[]{\label{fig:tower} 
  One of the, if not the, last NSO observing runs at the DST in the
  Sac Peak = NSO era. 
  Left-to-right: Alex Feller, Molchim Molnar, Lily Kromyda, Kevin
  Reardon, Valentin Mart{\'{\i}}nez Pillet, Francisco Iglesias,
  Franziska Zeuner.
  Always during my stays at Sac Peak I would walk up to the Tower in
  the morning to see what was happening. 
  During the Farewell a trio from MPS G\"ottingen were running,
  assisted by Kevin, a program on \SrI\ polarimetry; they are happy
  with their data.
  Here Valentin is touring with the two CU students that were the very
  last participant additions to the workshop, thanks to being driven
  by Kevin.  
  I would have welcomed many more: the very low influx of solar
  physicists coming out of CU (in fact most US universities) has been
  and still is a severe drawback to solar physics -- although
  fortunate for the many Europeans who found and find jobs in the US. 
  Even the splendid former Sac Peak summer-student program solicited
  more students from Freiburg than elsewhere.
  Hopefully, the strengthening of CU solar physics from the NSO
  relocation to Boulder will remedy this deficiency. 
  But maybe you should regard this statement as frustration of a
  member of a university, in fact a whole nation, that scrapped solar
  physics altogether -- presently all Dutch solar physicists except me
  work abroad, the Dutch Open Telescope stands mothballed even though
  its wide-field Halpha is better than anybody's, there is no hope of
  reviving Utrecht's former eminence in educating solar physicists. 
  In the 2015 solar publication statistics of
  \citetads{2016SoPh..291.1267S} 
  Holland ranks only 45th with Uganda, whereas there were more
  Farewell participants originally stemming from Utrecht than from any
  other education institute (the one solar physicist stemming from
  Uganda was Utrecht-educated too: Catherine Fischer now at Freiburg).  
  Sic transit gloria -- but the same may hold for this photograph.}
\end{figure} 

\paragraph*{\dropcap{F}arewell workshop.}

I was SOC co-chair and had proposed to have this workshop during a
2013 dinner in Cloudcroft. 
Sort of our farewell dinner after an extended stay; we planned the
Western Bar but it was closed for the yearly Christmas employee party
and so we ended up in a diner to the east with even crummier food and
no beer.
There I suggested it should be a fancy well-attended farewell and also
topic and title. 
Valentin reacted: ``OK, good idea, why don't you do it, but wait a few
years until DKIST gets closer to ready''. 
I first wanted it in 2015 and then again in 2016, fearing that the Sac
Peak infrastructure (notably Rex Hunter and Rebecca Coleman) would be
gone if we postponed too long, but those years got vetoed by Valentin. 
In the end we made it just before Sac Peak folded as an NSO site, so
perfectly though only just in time -- Rex and Rebecca were still there
and did their superb usual to my great relief. 

Rebecca even got me a hummingbird feeder, standing ready when we
arrived, and reminded us to bring a large quantity of sugar.  Viewing
hummingbird antics at our VOQ window (in 1977 it was the BOQ for
Bachelor Officer Quarters) was the highlight of this farewell for me.

At Steve's retirement home in Pagosa Springs (on a muddy dirt road
fittingly called ``Luxury Place'') I had told him in 2015, visiting on
our roundabout way to HAO after once again canoeing the White and
Green Rivers (Rangely to Sand Wash, our fourth descent), that I wanted
to be SOC boss, a joke alluding to the very many years in which his
email address was boss@nso.edu or such.   
He indeed gracefully gave me free reign in setting up the science part
of the program, basically gathering colleagues I like, while taking
care of the rest himself: applying for and getting NSF funding,
running the historical sessions, and overseeing LOC business.

I did the participant gathering not on my own but with a helpful SOC. 
The date had been shifted, from an earlier spring date with better
weather expectation but discarded due to overlap with the Seattle IRIS
workshop, to the week a week before the eclipse in order to permit
leisurely travel north to totality. 
Unfortunately, for quite some European colleagues the eclipse also
coincided with the start of the school season, while for some others
(notably Serge Koutchmy and Sara Martin) the workshop came too close to
their eclipse preparations, but the combination has probably enticed
some participants to travel to the US.

In addition to SolarNews announcements by Steve I sent personal
notifications to many colleagues whom the SOC helped define as
desirable.
Many responded enthusiastically.
So many that we feared that we might overrun the Visitor Center seat
capacity and therefore installed a pre-registration round with the
intention that the SOC should perform attendant selection if
necessary, not simply first = in.
In the end we did get close to the limit, but then late cancelations
(including four unwilling or unable to enter the US due to its new
president) outgrew late additions; in the end we managed fine.

The program went fine too, as evident from the talk pdf's on NSO's
meeting website; many excellent, only a few not up to par. 
As to posters: I have to confess I did not see any, using the breaks
instead to add 70 colleague portraits to my astronomer mugshot
collection (google ``Rob Rutten website'', descend to ``Photos''). 
Poster attendance is nearly always a problem at conferences; I hope my
colleagues did better than I did (while still getting their mug into
my camera).

\paragraph*{\dropcap{F}arewell ceremony.}
I had proposed the following choreography.
At the workshop's finish all participants walk up to the Tower and
line up along the visitor walkway around the telescope table. 
Then the three ex-directors present (Leibacher, Beckers, Keil)
representing the past spin the telescope up hand over hand, faster and
faster -- easy to do thanks to Dick Dunn's no-friction mercury float
hanging the immense 100-m 300-ton structure; our children and I used
to do this for fun on overcast Sundays -- and then, pivoting at the
center Valentin hands over the ceremonial keys to James, together at
the center of this revolving universe and so performing the
transition.
Applause and cheering from the event-horizon around, and then just as
much work to slow the telescope down again because the float is so
frictionless that it needs as much effort to spin it down as to spin
it up, again hand over hand but now by the seven students present
representing the future for DKIST and solar physics.
Thank-you-all by the NSO director, welcome-you-all by the new site
boss, festive toast.

It would have made a good YouTube memento -- but the two performers vetoed
it.
Instead, the meeting frazzled out.

\paragraph*{\dropcap{M}orals.}

I learned many things during this year, in solar physics, about
the US, and about myself.   
In solar physics at that time the highlight was the coming of age of
helioseismology. 
It was a very exciting development and I followed it closely at
Utrecht and kept up to date since but never worked in it; nor did I
appreciate that Tim Brown's invention then and there of his Fourier
tachometer was going to enable its maturing in the form of GONG, MDI,
HMI and local helioseismology.

The rest of solar physics remained pretty much in the infamous stage of
dermatology. 
Space data were virtually non-existent after the so promising Skylab
episode.
Numerical simulations hadn't taken off yet -- {\AA}ke Nordlund's
landmark $64\times64\times64$ granulation simulation got public in
Steve Keil's workshop in 1983. 
The mainstay was still my own field, spectroscopy and line formation. 
Avrett's VAL-I (1973) and VAL-II (1976) publications were out and already the
summit of numerical atmospheric modeling, but his monumental VAL-III
(1981) publication\footnote{\citetads{1981ApJS...45..635V}, 
the classic in describing the solar atmosphere as plane parallel
layers in hydrostatic and statistical equilibrium without dynamics or
magnetism. 
I love it deeply for my spectrum-formation teaching because every
spectral feature it produces (especially in its wonderful Figure 36) is
fully understandable, but nevertheless it is a severe mistake to take
it as representative of a mean solar atmosphere -- colleagues who
write about ``the temperature minimum'' or ``the transition region''
are plane-parallel-star physicists instead of solar physicists.
Both concepts do hold instantaneously along any microscopic radial column
through the solar atmosphere, but that's all -- certainly not 
plane-parallel layers or quasi-spherical shells in any sort of
equilibrium.  The Sun is more interesting than that!} still had to come.

Most observational solar physicists still sat squarely in the
photosphere and debated granules (George mesogranules but these seem to
have gone away) with fluxtubes on the horizon; only a few dared up
into the chromosphere. ``What about spicules?'' 
remained the nastiest question one could ask at any solar physics talk
(as at that time ``What about binaries?'' to stellar evolutionaries,
``What about magnetic fields?'' to galactic types, ``What about
neutrinos?'' to cosmologists. 
Nowadays ``What about dark matter'' and ``What about dark energy'' are
too easy since our neighbors ask those; naturally the answer is hydrogen
42, or maybe ice-nine).

The major improvements since then are threefold: the advent of
diffraction-limited solar imaging (the SST exemplary), the advent of
full-time many-diagnostic monitoring from space (SDO exemplary), the
advent of realistic simulation codes (Bifrost exemplary). 
The first development is essential because the Sun is intrinsically
small-scale in feeding mass and energy to its outer atmosphere.
The second is essential because (in Alan Title's words\footnote{In
Kiev in May 1989 at the only solar IAU symposium ever in the Soviet
Union, held at my instigation and ended by writing Gorbachov and Bush
to make them behave (\citeads{1990IAUS..138..529.}).
Guess what happened that autumn.})
the Sun is a complicated animal, needing holistic many-diagnostic,
high-resolution, large extent (space, time, spectral, Fourier)
observation; if you are observational but not co-aligning large-field
long-duration SDO, IRIS, Hinode, groundbased (and hopefully ALMA) data
sets your work may be besides the point.  
The third development says it is high time to stop interpretation
using static one-dimensional modeling -- far too last-century.

Radiative transfer and line formation are still important but no
longer the main topic. 
At present I find myself invited to teach it worldwide because
generally it has dropped out of the astronomy curriculum whereas a good
grasp is still needed for data interpretation if you are not
chianti-thin -- not too many colleagues are optically thick these
days\footnote{All who write that a particular line is thin are
thin themselves.  The width of a line is measured in \AA\ or Hz or \kms.
What they mean is that the feature they study appears transparent
in the line.}. 
But also for me it is now more a topic for teaching than for research
per s\'e. 
At Sac Peak I learned to ask what solar phenomena one diagnoses with
what lines, putting the science questioning and emphasis on the
former, knowledge to the latter.

Lessons about the USA: of course it differed much from what we
anticipated. 
We were never threatened by rednecks at gunpoint but found many people
amiable and hospitable. 
Most we appreciated the hugely wide empty expanses of New Mexico and
the spectacular scenery of the western deserts where we returned many
times since.   

Morals for me: I like writing computer programs but they bear fruit
too rarely: too often a wasted effort. 
Also the ones I developed at Sac Peak or took home (as Bruce Lites'
SOSO code which I much expanded but never used in publications, then
also an extensive Fourier package, idem).  
For some decades thereafter I didn't program at all because my
students were much better at it: I quit when we stopped having Control
Data computers serve whole universities at the time that Han
Uitenbroek and Jo Bruls started their PhD research and had to cope
with many operating systems in too fast succession.
Years later, at Mats Carlsson's prompting\footnote{Our collaboration
is another Sac Peak story. 
We met at Bruce Lites' 1984 workshop, sharing a relocatable where our
conversation naturally turned to ice-skating rivers and lakes and the
sea -- although all such were there as far away as they can get. 
End of February 1986 Mats visited Holland for skating, from our house
the frozen Linge river (the longest Dutch river originating in
Holland, our stretch navigable for barges up to 400~tons) and also
part of the route of the 200-km Eleven City Tour held a few days
before (and completed incognito by then crown-prince Willem Alexander
who also performed the DOT First Light ceremony in 1997), then joined
me a day at Sonnenborgh to make the visit chargeable, and told me to
learn TeX, emacs, and IDL. 
I learned all three (LaTeX rather then TeX) from Peppe Severino a year
later at Naples during a stay there, and have never left these as my
daily workhorses -- being very happy that they have been a stable
environment throughout the decades since.
Later Mats made me visiting professor at Oslo.
I went there many times and still visit Oslo regularly with much
pleasure.
At Utrecht my so-enhanced status helped gain extra years of support
for the Dutch Open Telescope and even led to temporary separation between
solar physics and astronomy, but I thought that a bad idea and undid
it later -- not knowing that astronomy would be killed altogether.}
I learned Latex (also a computer program and to be treated as such)
which served me well to co-write (or write) student publications, but Unix
and IDL took me much longer because my students did all the tricks
including maintaining my home computer: for a decade an Atari-ST and a
beloved ST-book that I took to Sac Peak and La Palma, one of the few
hundred ever made; then laptops from Dell and Toshiba running linux,
then an unhappy MacBook episode when Alfred de Wijn and Jorrit
Leenaarts decided that I should do without further help from them and
could do no harm to a Mac -- but I went back to linux at my discovery
that googling helpful Stackexchange nowadays replaces helpful
students.
Inventorying colleagues at meetings shows that presently 90\% laps a
Mac but I proudly do not, instead belonging to the happy-few elite of
post-Mac solar scientists.
I also slowly stumbled my own way into IDL. 
Some of my programs actually became productive; my ``simple IDL
manual'' for dummies like myself is now one of my most downloaded
documents and I hope that the SDO co-alignment pipeline that I spent
much of the past year developing will be useful to others too.
I still hesitate re Python, although I have web-found my IDL manual
translated line-by-line into a Python manual. 

I also liked and still like observing, but my track record in reducing
data to get fit for analysis and publication has remained dismal. 
It turned out much much better to rely on colleagues to get me
fully-reduced data to analyze and publish: Bruce Lites, Kevin Reardon,
in recent years especially Luc Rouppe van der Voort who is very good
at feeding me what I want or should want\footnote{Luc is also my first
offspring outhirsching me.
Sign of a field in good shape: Title/Leighton = 2.2, Linsky/Avrett =
2.1, Schrijver/Zwaan = 2.0, Carlsson/Scharmer = 1.6, Lites/Athay =
1.5, Solanki/Stenflo = 1.5, Schmieder/Mein = 1.4, Nordlund/Gustafsson
= 1.2, and other pupil/teacher combos. 
But this ratio seems declining: is solar physics over the top?}. 
This dependence/reliance on others started at Sac Peak.

My best lesson was that I much enjoyed collaborating with colleagues
and that such collaborations were and are fruitful because they push
me over the psychological barrier to sit down and write.
Without some personal obligation I find that hard to do; only recently
have I published single-author journal publications since my thesis ones
(Gene Parker has hundreds! 
He must be the champion single-author hirscher). 
The happy start of my collaborative publishing was also at Sac Peak.

Later having graduate students forced me to come up with viable
research suggestions beyond the endless tinkering I tend to get stuck
in myself.   
They were not only useful to my self-esteem by showing myself that I
could inspire them and so get results (I am proud of my scientific
offspring -- five at the Farewell\footnote{In PhD date order: Han
Uitenbroek, Luc Rouppe van der Voort who wasn't my graduate student
but I did tweak his nose forever to La Palma, Alfred de Wijn, Jorrit
Leenaarts, Gregal Vissers who got his PhD with Luc but I also tweaked
his nose, and additionally Michiel van Noort whose career I tweaked
only during minutes but decisively, and Mandy Hagenaar who visited the
workshop from her summer cabin in nearby Timberon. 
All my solar offspring works abroad; in Holland Thijs Krijger runs a
satellite-design consultancy, Frans Snik moved with Christoph Keller
to exoplanets.}) but also in boosting my own publication production,
although it never rose beyond 1-2/year. 
This phase of tutoring youngsters is now over, much to my regret, but
I am grateful to colleagues still willing to share data and research
with me and inviting me to teach.

{\footnotesize
\bibsep=0ex
\bibliographystyle{aa-note} 
\bibliography{rjrfiles,adsfiles} 
}

\end{document}